\documentclass[a4paper,fleqn,usenatbib]{mnras}

\usepackage{newtxtext,newtxmath}

\usepackage[T1]{fontenc}
\usepackage{ae,aecompl}

\usepackage{graphicx}	
\usepackage{amsmath}	
\usepackage{amssymb}	
\usepackage{times}
\usepackage{xspace}

\usepackage{color}
\usepackage{hyperref}
\usepackage{supertabular}
\hypersetup{colorlinks, linkcolor=blue}

\def\msun{{\rm M}_{\sun}}

\title[The core of MACS\,J0417 with MUSE]{The core of the massive cluster merger MACS\,J0417.5-1154 as seen by VLT/MUSE}

\author[Jauzac et al. 2018]{
Mathilde Jauzac,$^{1,2,3}$\thanks{E-mail: mn@ras.org.uk (KTS)}
Guillaume Mahler,$^{4}$
Alastair C. Edge,$^{1}$
Keren Sharon,$^{4}$
Steven 
\newauthor Gillman,$^{1}$
Harald Ebeling,$^{5}$
David Harvey,$^{6}$
Johan Richard,$^{7}$
Stephen L. Hamer,$^{8}$
\newauthor
Michele Fumagalli,$^{1,2}$
A. Mark Swinbank,$^{1}$
Jean-Paul Kneib,$^{5}$
Richard Massey,$^{1,2}$
Philippe 
\newauthor Salom\'e$^{9}$
\\
\\
\\
$^{1}$Centre for Extragalactic Astronomy, Department of Physics, Durham University, Durham DH1 3LE, UK\\
$^{2}$Institute for Computational Cosmology, Durham University, South Road, Durham DH1 3LE, UK\\
$^{3}$Astrophysics and Cosmology Research Unit, School of Mathematical Sciences, University of KwaZulu-Natal, Durban 4041, South Africa\\
$^{4}$Department of Astronomy, University of Michigan, 1085 South University Ave, Ann Arbor, MI 48109, USA\\
$^{5}$Institute for Astronomy, University of Hawaii, 2680 Woodlawn Drive, Honolulu, Hawaii 96822, USA\\
$^{6}$Laboratoire d'Astrophysique, Ecole Polytechnique F\'ed\'erale de Lausanne (EPFL), Observatoire de Sauverny, CH-1290 Versoix, Switzerland\\
$^{7}$CRAL, Observatoire de Lyon, Universit\'e Lyon 1, 9 Avenue Ch. Andr\'e, 69561 Saint Genis Laval Cedex, France\\
$^{8}$Institute of Astronomy, University of Cambridge, Madingley Road, Cambridge CB1 0HA, UK \\
$^{9}$LERMA, Observatoire de Paris, CNRS, PSL University, Sorbonne University, 75014 Paris, France
}

\date{Accepted XXX. Received YYY; in original form 2018}

\pubyear{2018}

\begin{document}
\label{firstpage}
\pagerange{\pageref{firstpage}--\pageref{lastpage}}
\maketitle

\begin{abstract}
We present a multi-wavelength analysis of the core of the massive galaxy cluster MACS\,J0417.5-1154 ($z = 0.441$). Our analysis takes advantage of VLT/MUSE observations which allow the spectroscopic confirmation of three strongly-lensed systems. 
System \#1, nick-named \emph{The Doughnut}, consists of three images of a complex ring galaxy at $z = 0.8718$ and a fourth, partial and radial image close to the Brightest Cluster Galaxy (BCG) only discernible thanks to its strong [OII] line emission. 
The best-fit mass model (rms of 0.38\arcsec) yields a two-dimensional enclosed mass of $M({\rm R < 200\,kpc}) = (1.77\pm0.03)\times10^{14}\,\msun$ and almost perfect alignment between the peaks of the BCG light and the dark matter of ($0.5\pm0.5$)\arcsec. 
We observe a significant misalignment when system \#1 radial image is omitted. The result serves as an important caveat for studies of BCG-dark matter offsets in galaxy clusters. 
Using \emph{Chandra} to map the intra-cluster gas, we observe an offset between gas and dark-matter of ($1.7\pm0.5$)\arcsec, and excellent alignment of the X-ray peak with the location of optical emission line associated with the BCG. 
We interpret all observational evidences in the framework of on-going cluster merger activity, noting specifically that the coincidence between the gas and optical line peaks may be evidence of dense, cold gas cooled directly from the intra-cluster gas. 
Finally we measure the surface area, $\sigma_{\mu}$, above a given magnification factor $\mu$, a metric to estimate the lensing power of a lens, $\sigma(\mu > 3) = 0.22$\,arcmin$^2$, which confirms MACS\,J0417 as an efficient gravitational lens.

\end{abstract}




\section{Introduction}
Galaxy clusters are the most massive gravitationally bound objects in the observable Universe and are connected to each other on large scales via a web of filaments of matter, commonly referred to as the cosmic web \citep[e.g.][]{bond96}. 
The location of galaxy clusters in this web makes them undergo constant growth through mergers with smaller clusters and groups of galaxies and surrounding material \citep{springel05,schaye15,colless01}. This non-equilibrium state allows us to witness evolution and formation processes \citep{dietrich12,jauzac12,jauzac16b,jauzac18b} that are not readily observable in other laboratories.
Among the many aspects of modern astrophysics and cosmology addressed by cluster studies are the nature of dark matter and its physical properties, galaxy evolution in dense environments, and the influence of active galactic nuclei (AGN) on their host galaxy and local environment, all of which yield observables that play an important role in the calibration of the next generation of cosmological hydrodynamical simulations such as \textsc{Eagle} and \textsc{Illustris} \citep{schaye15,vogelsberger14}.  

Due to their high mass, clusters of galaxies also act as gravitational telescopes. They deflect light emitted by galaxies behind them, creating magnified and distorted images. This gravitational lensing effect is the most powerful tool known to map the total matter in clusters (both dark and luminous), as it is purely geometrical and independent of the dynamical status of the cluster lens \citep[for reviews see e.g. ][]{bible,massey10,KN11,hoekstra13}.
Gravitational lensing by clusters has become a widely used tool in the course of the past decade. In addition to revealing fundamental properties of dark matter, lensing magnifies the light emitted by background galaxies, and thus can extend telescope's observational reach by up to six magnitudes compared to the (unlensed) fields and thus enables study of the high-redshift Universe \citep[as an example among others, see][]{atek15b,atek18,bouwens17a,bouwens17b,ishigaki15,ishigaki18,kawamata18,livermore17,kawamata18}. For a review on galaxy evolution, we refer the reader to \cite{dayal18}. A significant amount of observing time with the \emph{Hubble Space Telescope} (HST) had been invested in observations of galaxy cluster lenses, e.g., the MAssive Cluster Survey SNAPshot programs \citep[MACS, PI: Ebeling,][]{ebeling01}; the Cluster Lenses And Supernovae with Hubble Treasury programme \citep[CLASH, PI: Postman,][]{postman12}; the SDSS Giant Arcs Survey (SGAS, PI: Gladders, Sharon et al.\ \emph{in prep.}); the Grism Lens-Amplified Survey from Space \citep[GLASS, PI: Treu,][]{schmidt14}; the \emph{Hubble Frontier Fields} \citep[HFF, PI: Lotz,][]{lotz17} which collected the deepest data ever on six galaxy clusters; the REionization LensIng Cluster Survey \citep[RELICS, PI: Coe,][]{salmon17}; and since July 2018 the Beyond the Ultra-deep Frontier Fields And Legacy Observation programme (BUFFALO, PIs: Steinhardt \& Jauzac, Steinhardt et al.\ in prep.) which will observe the outskirts of the HFF clusters.

Observations of, in particular, cluster cores have also greatly benefited from the Multi-Unit Spectroscopic Explorer \citep[MUSE,][]{bacon10} on the Very Large Telescope (VLT). MUSE is an optical IFU with a field of view of $1\times1$\,arcmin$^2$, well matched to the size of the high-magnification region in cluster cores (i.e., clusters featuring Einstein radii of 15-30\arcsec). MUSE observations provide information on the cluster galaxy population, as well as on foreground galaxies and lensed background populations \citep[e.g. ][]{richard15,grillo15,karman15,jauzac16a,treu16,caminha17a, caminha17b,annunziatella17,monna17,bonamigo18,chirivi18}.

We here present a multiwavelegth analysis of the core of MACS\,J0417.5-1154 (hereafter MACS\,J0417), a very X-ray luminous cluster at $z=0.441$ discovered in the course of the MACS survey \citep{ebeling01,ebeling10}. MACS\,J0417 is the second most luminous X-ray cluster in the MACS sample at $0.3<z<0.5$, and is classified as a binary, head-on cluster merger proceeding along an axis misaligned with our line of sight \citep{ebeling10,ME12}. The X-ray properties of the cluster and its dynamical status were also studied recently by \cite{botteon18} using the available \emph{Chandra} data.
The Brightest Cluster Galaxy (BCG) of MACS\,J0417 shows strong optical emission lines and atypical photometric colours, consistent with a classification of  `active' in the framework advanced by \citet{green16}, reinforcing the status of MACS\,J0417 as a dynamically evolving galaxy cluster.
At radio wavelengths, MACS\,J0417 was found to host a peculiar radio halo first discussed in \cite{dwarakanath11}, and more recently in \cite{parekh17} and \cite{sandhu18}. The radio diffuse emission is extended along the North-West direction, sign of a high velocity merger. The weak-lensing properties of MACS\,J0417 were first studied as part of the Weight the Giants project \citep[WtG,][]{vonderlinden14a}. \cite{applegate14}, who also describe the WtG weak-lensing mass measurement technique, cite a mass for MACS\,J0417 of $M(R<1.5\,{\rm Mpc})=1.9\times10^{15}\msun$, making MACS\,J0417 the fourth most massive cluster of their sample of 51 clusters. Recently, \cite{pandge18} presented a multi-wavelength anaylisis of the cluster combining X-ray observations from \emph{Chandra}, radio data from the Giant Metrewave Radio Telescope (GMRT) and Bolocam, and optical imaging from both Subaru and \emph{HST}. They confirm the merging status of MACS\,J0417, and provide a weak-lensing mass of the cluster, $M(R<1.9\,{\rm Mpc}=(1.4\pm 0.3)\times10^{15}\,\msun$. This estimate is slightly lower than what was measured by WtG, however remains of the same order of magnitude.

We here present a fresh view of MACS\,J0417's central strong-lensing region based on a VLT/MUSE observation obtained as part of a larger survey of massive cluster cores. To aid the interpretation of the MUSE results, we complement our analysis with \emph{HST}, \emph{Chandra} and VLT/SINFONI data. As MACS\,J0417 was also selected as one of the targets of the HST RELICS survey, an independent analysis of the system's core based on the RELICS data, and focusing on lensed, high-redshift galaxies, is presented in a companion paper by \cite{mahler18b}.

The paper is organized as follows: Sect.~\ref{sec:obs} describes the observations used in this work, Sect.~\ref{sec:slmodel} presents our strong-lensing mass model of the cluster, Sect.~\ref{sec:discussion} discusses implications of our findings for the dark-matter distribution of the cluster compared to the distribution of light and gas, the impact of assumptions regarding the redshift of the lensed galaxies, and differences between our model and the one presented in the companion paper by \cite{mahler18b}. We offer conclusions in Sect.~\ref{sec:conclusion}.

Throughout this paper, we adopt the cold dark matter concordance cosmology, $\Lambda$CDM, with  $\Omega_{m} = 0.3$, $\Omega_{\Lambda}= 0.7$, and a Hubble constant $H_{0}= 70$\,km.s$^{-1}$\,Mpc$^{-1}$. All magnitudes are quoted in the AB system. At the redshift of MACS\,J0417, an angular separation of 1\arcsec\ corresponds to 5.708\,kpc.

\begin{table}
\begin{center}
\caption{Catalogue of foreground objects detected in the VLT/MUSE data.}
\begin{tabular}[h!]{cccc}
\hline
\hline
\noalign{\smallskip}
ID & R.A. & Dec. & $z_{\rm spec}$\\
 & ($deg$) & ($deg$) & \\
\hline
\hline
F1 & 64.3895834 & -11.9112644 & 0.3471 \\
F2 & 64.3900500 & -11.9118044 & 0.2307 \\
\hline
\hline
\end{tabular}
\label{tab:fgd_muse}
\end{center}
\end{table}

\begin{table}
\begin{center}
\caption{Catalogue of cluster members detected in the VLT/MUSE data.}
\begin{tabular}[h!]{cccc}
\hline
\hline
\noalign{\smallskip}
ID & R.A. & Dec. & $z_{\rm spec}$\\
 & ($deg$) & ($deg$) & \\
\hline
\hline
BCG & 64.39456700 & -11.90893200 & 0.4408 \\
G1  & 64.38716272 & -11.90694454 & 0.4510 \\
G2  & 64.38780147 & -11.91254854 & 0.4485 \\
G3  & 64.38914780 & -11.90911027 & 0.4485 \\
G4  & 64.39357446 & -11.90701162 & 0.4407 \\
G5  & 64.39381502 & -11.91471127 & 0.4375 \\
G6  & 64.39430461 & -11.91000339 & 0.4380 \\
G7  & 64.39571928 & -11.90020200 & 0.4452 \\
G8  & 64.39676898 & -11.91493756 & 0.4385 \\
G9  & 64.39902786 & -11.90705624 & 0.4420 \\
G10 & 64.39932440 & -11.90180948 & 0.4180 \\
G11 & 64.39991770 & -11.90453332 & 0.4340 \\
G12 & 64.40110432 & -11.90984703 & 0.4354 \\
G13 & 64.40185734 & -11.91090753 & 0.4490 \\
G14 & 64.40274724 & -11.91098565 & 0.4470 \\

\hline
\hline
\end{tabular}
\label{tab:cm_muse}
\end{center}
\end{table}

\section{Observations}
\label{sec:obs}
\subsection{VLT/MUSE}
\begin{table}
\begin{center}
\caption{Catalogue of singly imaged background galaxies detected in the VLT/MUSE data.}
\begin{tabular}[h!]{cccc}
\hline
\hline
\noalign{\smallskip}
ID & R.A. & Dec. & $z_{\rm spec}$\\
 & ($deg$) & ($deg$) & \\
\hline
\hline
\noalign{\smallskip}
B1   & 64.40005721 & -11.90873165 & 0.5635 \\
B2   & 64.39423617 & -11.90178719 & 0.5820 \\
B3   & 64.39079058 & -11.91616551 & 1.0560 \\
B4   & 64.38910205 & -11.91475891 & 0.8089 \\
B5   & 64.39975818 & -11.91643341 & 1.0457 \\
B6   & 64.39521935 & -11.90464675 & 0.5026 \\
B7   & 64.40075417 & -11.90402222 & 3.7450 \\
B8   & 64.39584583 & -11.91195889 & 0.5023 \\

\hline
\hline
\end{tabular}
\label{tab:bkg_muse}
\end{center}
\end{table}

The integral field spectrograph MUSE \citep{bacon10} observed the core of MACS\,J0417 as part of a larger survey of MACS clusters with MUSE using poorer weather conditions (ESO project 0100.A-0792(A), PI: Edge). The observation was taken on 2017 December 12 in clear conditions but with strongly varying seeing between 1.3\arcsec and 1.9\arcsec. The exposure was split in 3$\times$970\,s, resulting in 2.91\,ks in total.

The MUSE data were reduced with the ESO pipeline \citep[][v2.4]{weilbacher15} and then reprocessed to improve flat-fielding and sky subtraction using a dedicated pipeline together with \textsc{mpdaf} tools \citep{bacon16}, following a methodology similar to the one described in \cite{fumagalli16,fumagalli17}. We refer the reader to these papers for more details and only give a brief summary of the post-processing steps here: (1) a resampling of the individual exposures is performed to a common astrometric grid defined by the final ESO reduction; (2) a correction of imperfections in the flat-fields is applied using the \textsc{mpdaf} self-calibration method as described in \cite{bacon17}, and sky subtraction is performed with the \textsc{zap} tool which uses principal-component analysis \citep{soto16}; (3) individual exposure are combined to obtain a single data cube using an average 3$\sigma$ clipping algorithm.

The detailed analysis of the final data cube allowed us to extract spectra of 36 sources that we classified into foreground objects, cluster members, singly imaged background objects, and multiply imaged objects. Tables~\ref{tab:fgd_muse}, \ref{tab:cm_muse} and \ref{tab:bkg_muse} list foreground, cluster, and singly imaged background objects respectively. 
The multiple images used in this work are listed in Table~\ref{tab:immul}, and the redshift of the ones that have been spectroscopically confirmed are given in the $z$ column without error bars. 
The multiple image set used in this work is discussed in more detail in Sect.~\ref{sec:immul}. 

Other serendipitous discoveries are presented in the Appendix of this paper. These include a galaxy triplet in the background of the cluster at $z=1.046$ (Appendix~\ref{sec:galpair}) and an extremely dense starburst at $z=0.56$ (Appendix~\ref{sec:greenpea}).

\begin{figure*}
\begin{center}
\includegraphics[width=0.815\textwidth]{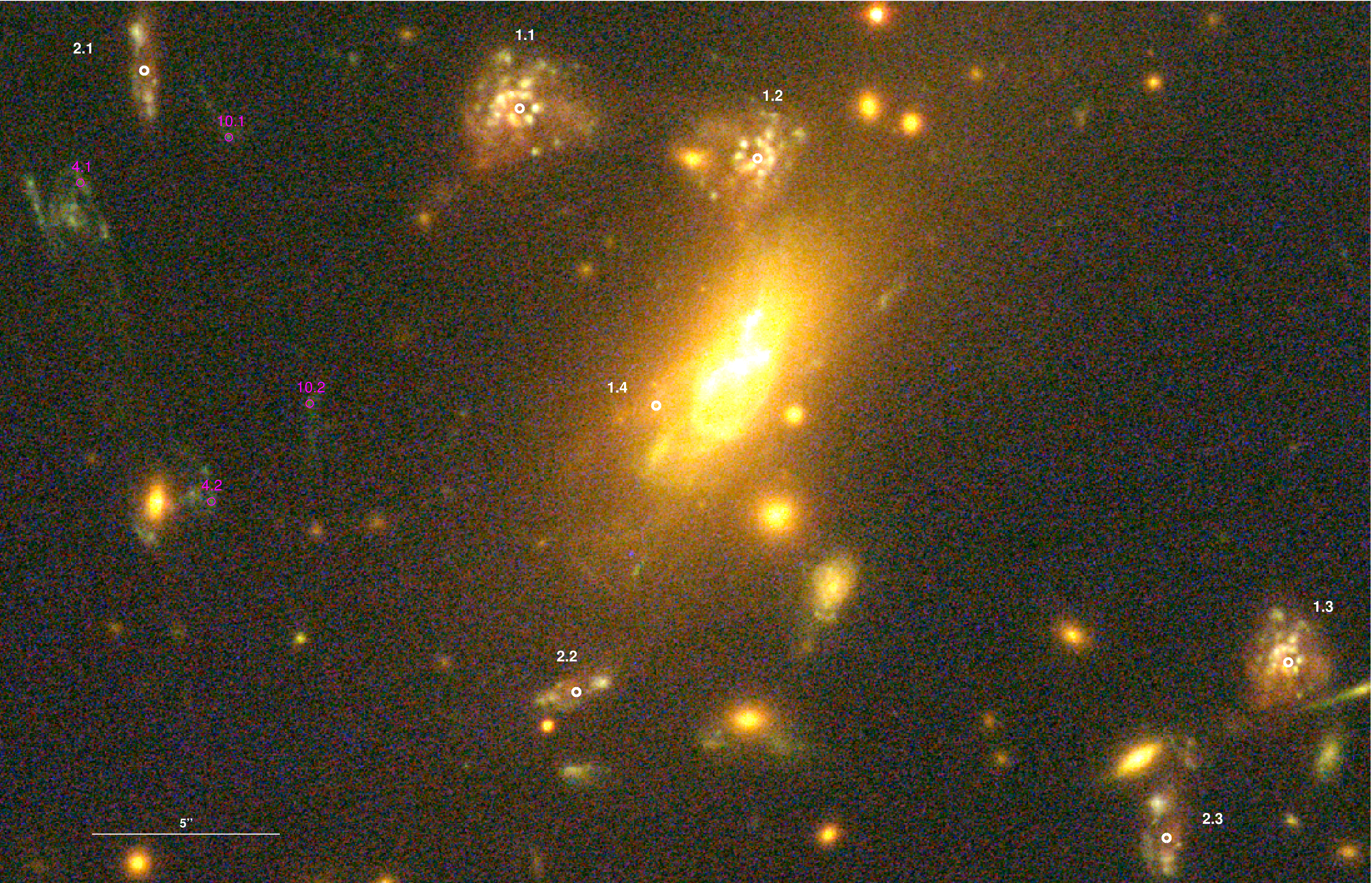}\
\includegraphics[width=0.2\textwidth,angle=0.0]{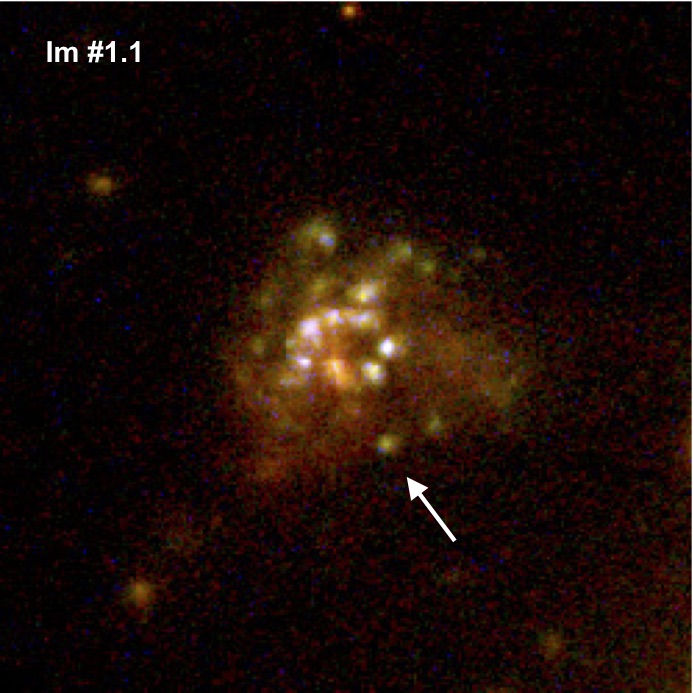}\
\includegraphics[width=0.2\textwidth,angle=0.0]{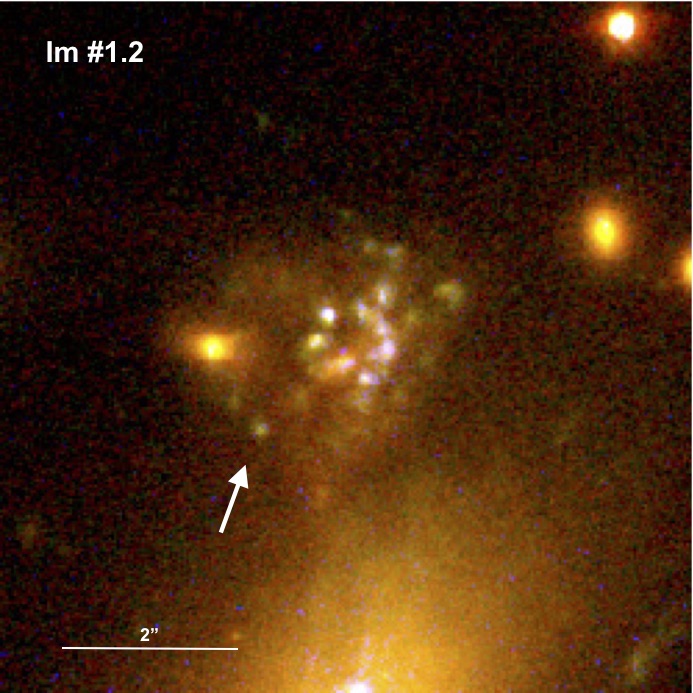}\
\includegraphics[width=0.2\textwidth,angle=0.0]{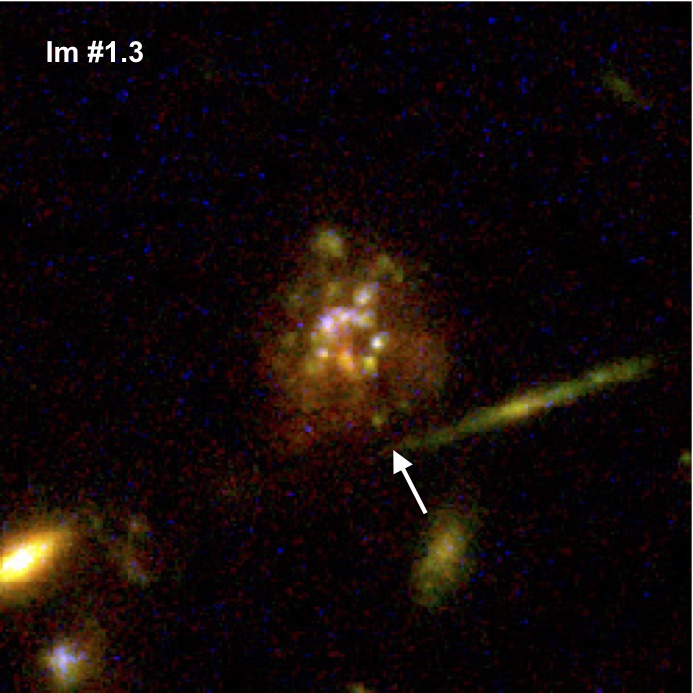}\
\includegraphics[width=0.2\textwidth,angle=0.0]{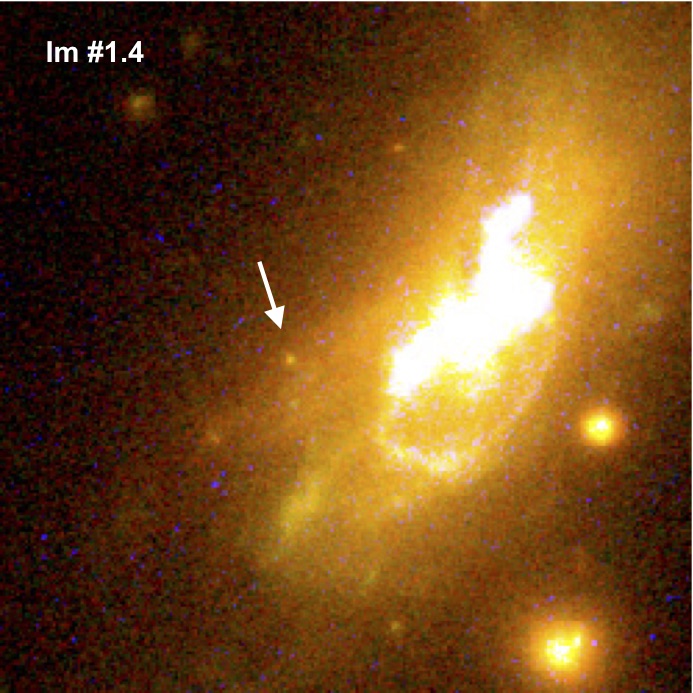}\ 
\\
\includegraphics[width=0.2\textwidth,angle=0.0]{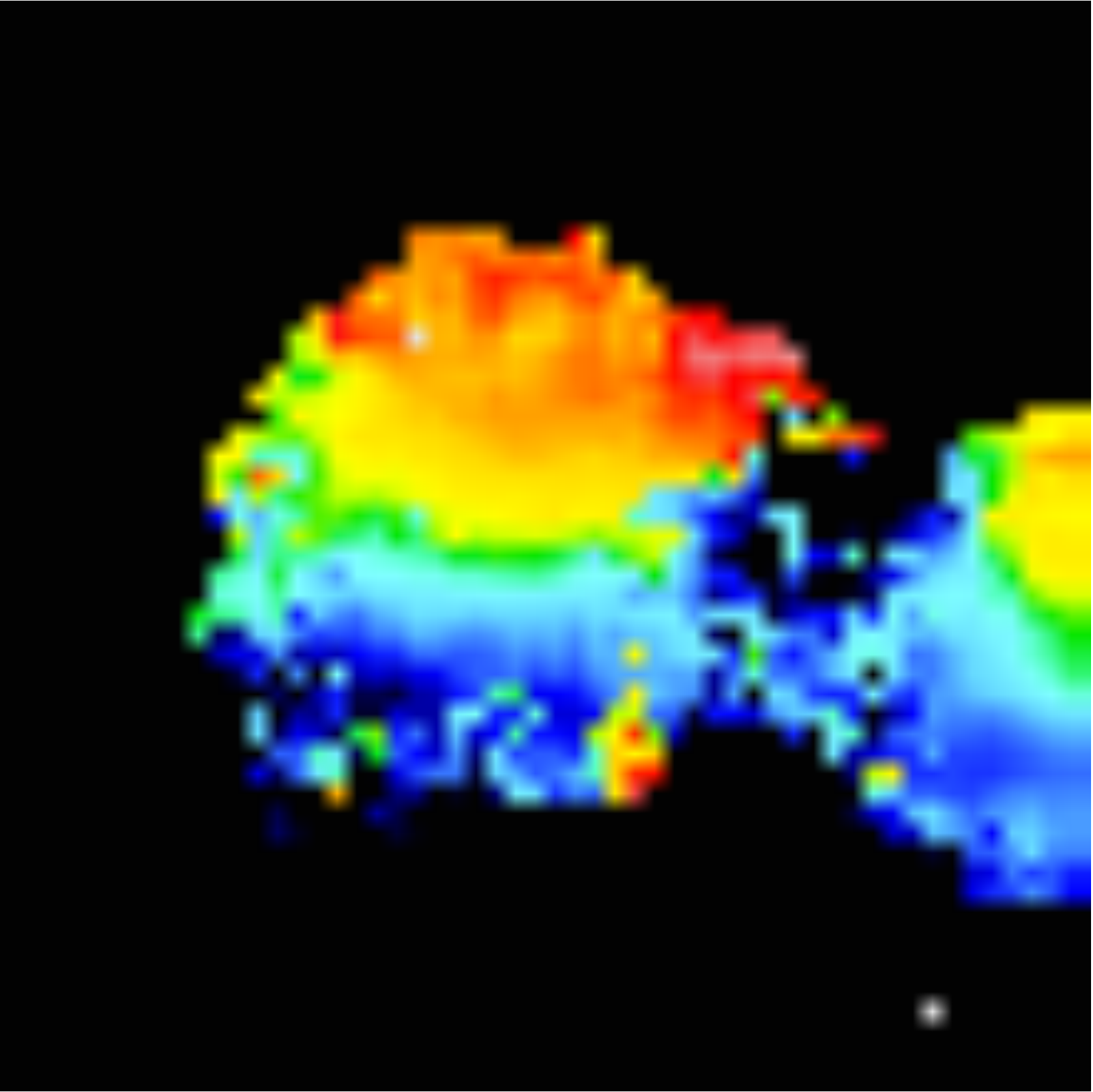}\ 
\includegraphics[width=0.2\textwidth,angle=0.0]{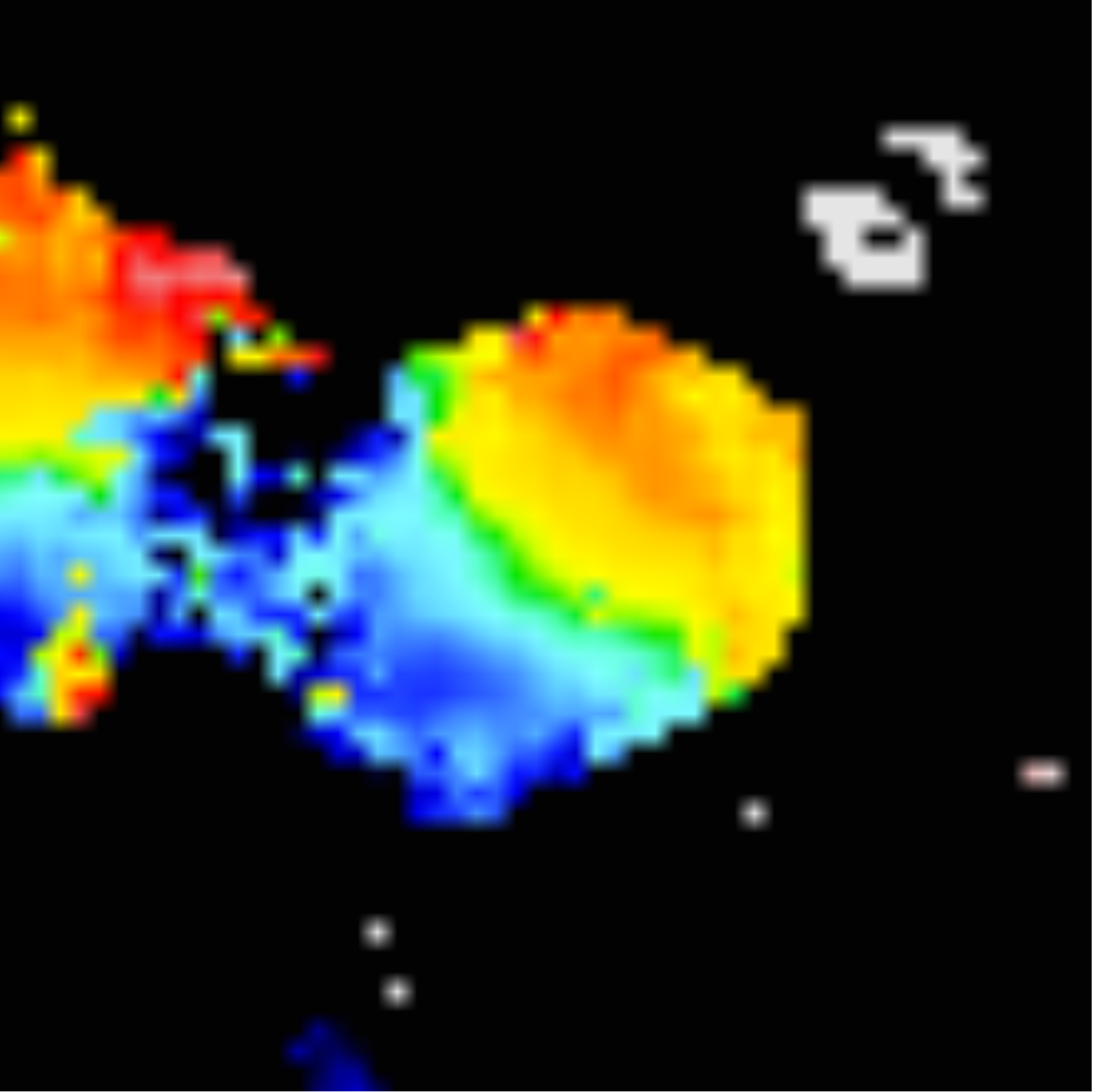}\ 
\includegraphics[width=0.2\textwidth,angle=0.0]{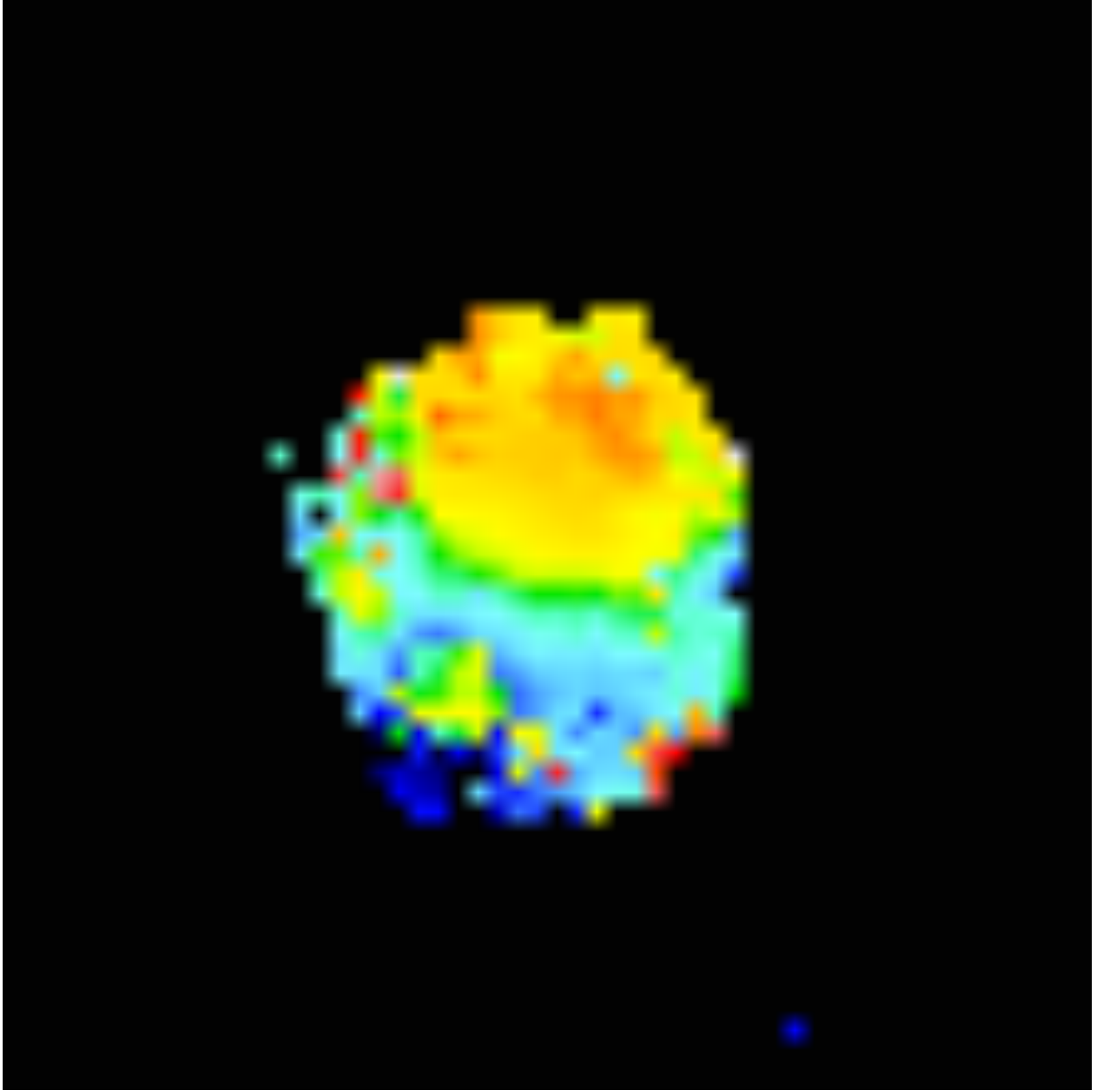}\ 
\includegraphics[width=0.2\textwidth,angle=0.0]{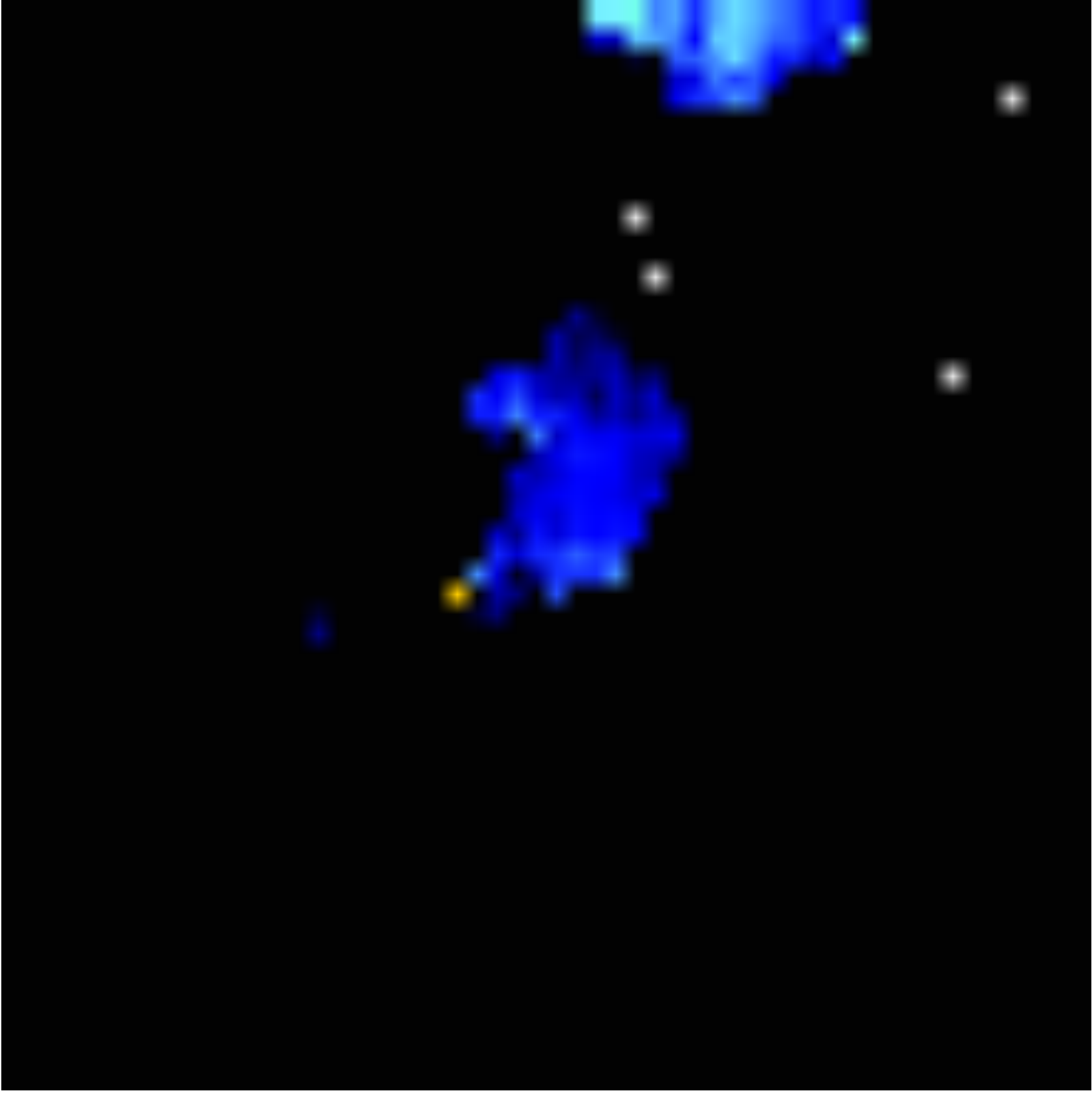}\ 
\\
\includegraphics[width=0.2\textwidth,angle=0.0]{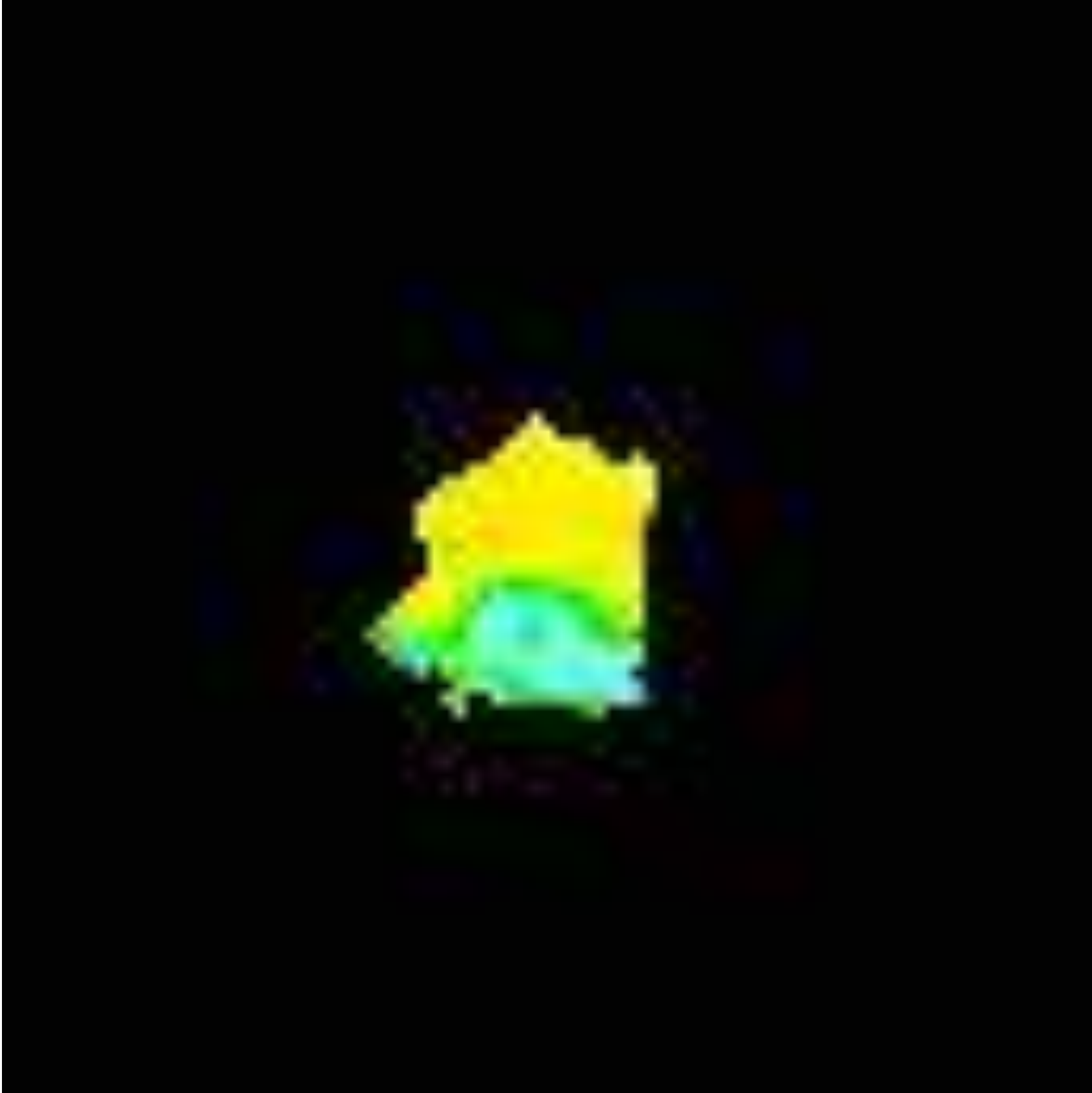}\ 
\includegraphics[width=0.2\textwidth,angle=0.0]{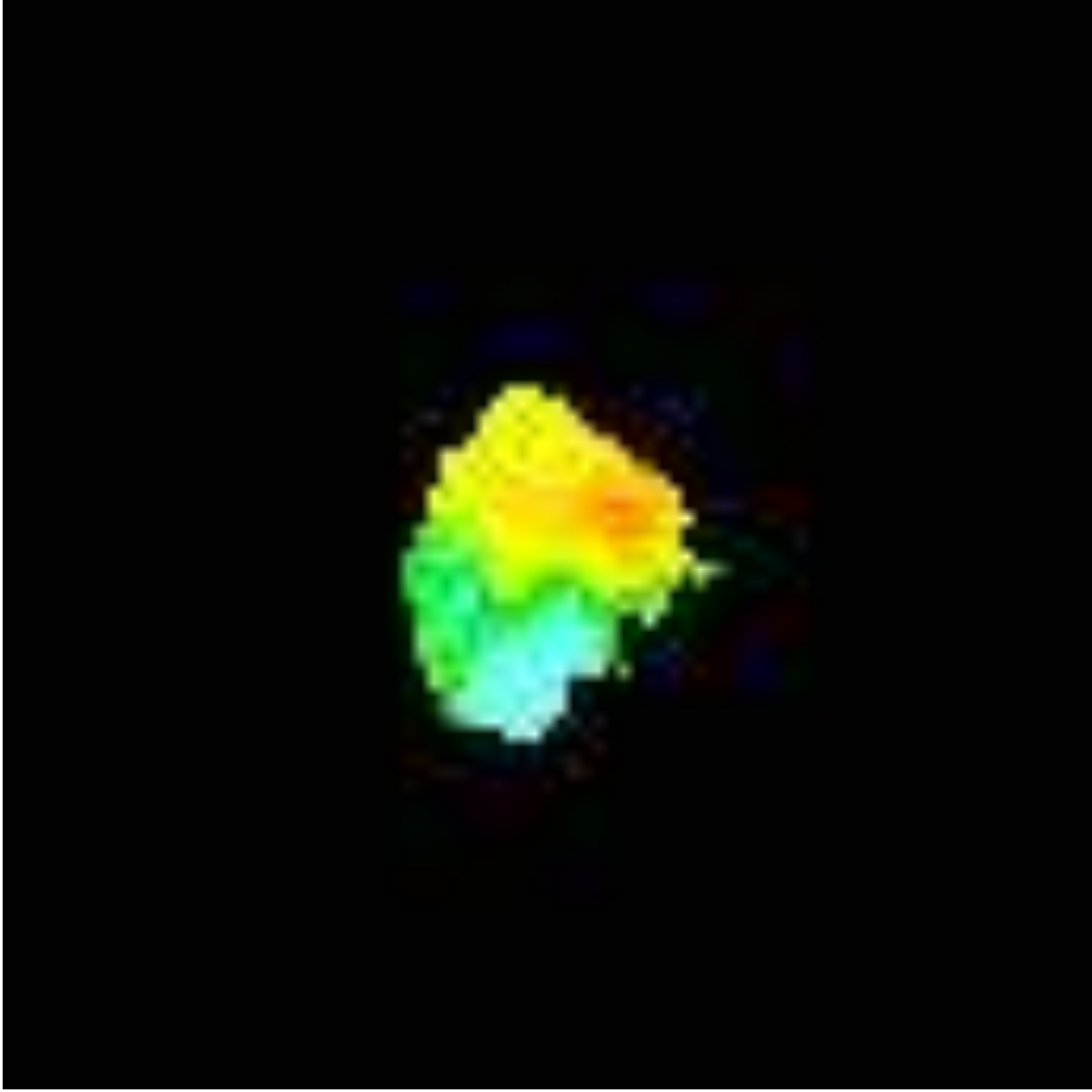}\ 
\includegraphics[width=0.2\textwidth,angle=0.0]{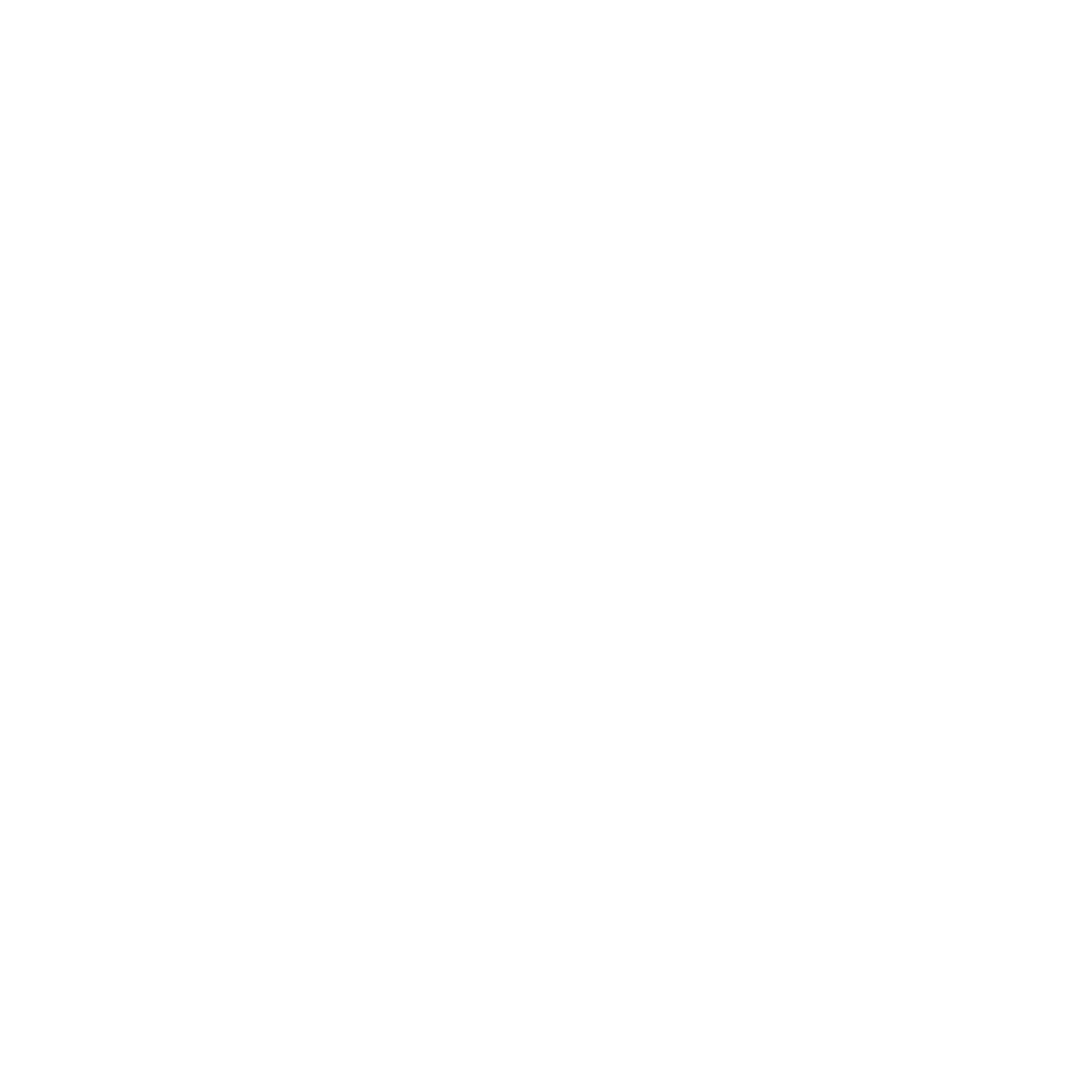}\ 
\includegraphics[width=0.2\textwidth,angle=0.0]{empty.pdf}\ 
\\
\caption{\emph{Top panel:} Zoom into the core of MACS\,J0417. We here show a colour composite image from \emph{HST}/ACS and WFC3/UVIS images. The white circles highlight the multiple images of systems \#1 and \#2 spectroscopically confirmed by our VLT/MUSE observations. North is up, East if left, and magenta circles show the other multiply-imaged systems visible in this field of view.
\emph{Second panel:} \emph{HST} colour composite stamps of all four images of system \#1, also nicknamed \emph{The Doughnut}. We highlight with an arrow the part of \emph{The Doughnut} that is quadruply-imaged.
\emph{Third panel:} MUSE low-resolution [OII] velocity maps of the four images of system \#1. The stamps are the same size
as the \emph{HST} ones and cover a velocity range of -1500 to +1500\,km.s$^{-1}$ and are fit above a threshold of 6$\sigma$ in bins averaged over
0.6\arcsec in size. Note the line emission of other images appear in the margin of each image apart from \#1.3 which is the most isolated. 
\emph{Bottom panel:} The velocity maps extracted from fitting the SINFONI spectra on the spaxel scale. The 
range in velocity is consistent with that recovered in the SINFONI observation when the poorer surface brightness sensitivity in the NIR is considered.
The scale given on the \emph{HST} stamp of image \#1.2 is applicable to all other \emph{HST}, MUSE, and SINFONI stamps.
}
\label{fig:donut}
\end{center}
\end{figure*}

\subsection{\emph{Hubble Space Telescope}}
\label{sec:obshst}
MACS\,J0417 was first observed with \emph{HST} using the \emph{Wide-Field Planetary Camera 2} (WFPC2) in 2008 as part of the MACS SNAPshot programme G0-11103 (PI: Ebeling) in the F814W and F606W passbands .
It was then observed with the \emph{Advanced Camera for Survey} (ACS) in the F814W passband, and with the \emph{Wide Field Camera 3} (WFC3) through its UVIS channels in the F606W passband in 2010 (PI: von der Linden, G0-12009).
It was subsequently observed in 2015 with ACS in the F435W passband, and with WFC3 in the F105W, F125W, 140W and F160W passbands as part of the RELICS programme (PI: Coe, GO-14096). Table~\ref{tab:hstobs} gives a summary of the \emph{HST} observations available for MACS\,J0417.

For the analysis presented here, we make use of the publicly available RELICS data products.\footnote{\url{https://archive.stsci.edu/missions/hlsp/relics/macs0417-11/images/}}
Observations with ACS and WFC3 were aligned and combined following the procedure described by \cite{cerny18}; we used both 0.03\arcsec and 0.06\arcsec pixel-size images. The top panel of Fig.~\ref{fig:donut} shows a zoomed view of the central region of MACS\,J0417 as observed by \emph{HST}; highlighted in white are two lensed systems spectroscopically confirmed by VLT/MUSE as listed in Table~\ref{tab:immul}.

\begin{table}
\begin{center}
\caption{Overview of the different \emph{HST} observations of MACS\,J0417 as described in Sect.~\ref{sec:obshst} including exposure time and date of observations for each filter and observing programme.}
\begin{tabular}[h!]{cccc}
\hline
\hline
\noalign{\smallskip}
Camera/Filter & Exp. Time & Date & Prog. ID\\
 & ($s$) &  & \\
\hline
\hline
WFPC2/F814W & 1200 & 2007-11-21 & 11103 \\
WFPC2/F606W & 1200 & 2007-11-21 & 11103 \\
ACS/F814W & 1910 & 2010-12-10 & 12009\\
WFC3-UVIS/F606W & 5364 & 2011-01-20 & 12009\\
WFC3-UVIS/F606W & 1788 & 2011-02-28 & 12009\\
ACS/F435W & 2000 & 2016-11-30 & 14096\\
WFC3/F105W & 356 & 2016-02-11 & 14096\\
WFC3/F105W & 756 & 2017-02-10 & 14096\\
WFC3/F125W & 381 & 2016-12-30 & 14096\\
WFC3/F125W & 356 & 2017-02-11 & 14096\\
WFC3/F140W & 381 & 2016-12-30 & 14096\\
WFC3/F140W & 356 & 2017-02-10 & 14096\\
WFC3/F160W & 1006 & 2016-12-30 & 14096\\
WFC3/F160W & 1006 & 2017-02-11 & 14096\\
\hline
\hline
\end{tabular}
\label{tab:hstobs}
\end{center}
\end{table}

\subsection{\emph{Chandra} X-ray Observatory}
MACS\,J0417 was observed by the \emph{Chandra X-ray Observatory} on three occasions (OBSID 3270, 11759, and 12010), for a total exposure time of 81.5\,ks. All observations were performed with ACIS-I. We retrieved the data directly from the \emph{Chandra} archive, as reduced with the \textsc{ciao} v4.10 pipeline. We did not reprocess the data and used the full-band images with no energy filter applied.

\subsection{VLT/SINFONI}
An archival VLT/SINFONI observation that covers two of the multiple images of system \#1 (1.1 and 1.2 as listed in Table~\ref{tab:immul}) was obtained as part of a survey of six lensed $0.7<z<1.1$ galaxies (087.A-0700(A), PI Limousin) on 2011 September 10 with the $J$-band grating and largest field mode of 8.0\arcsec $\times$ 8.0\arcsec. The spectral resolution in the $J$-band band is $\lambda / \Delta \lambda \sim$4500. Each observing block (OB) was taken in an ABBA observing pattern (A=Object frame, B=Sky frame)  with chops to sky to allow improved sky subtraction. 
The target was observed for a total of 2.7\,ks comprising nine individual exposures of 300\,s in $\sim$0.6\arcsec seeing and photometric conditions.
The observations were dithered and nodded to a sky position. 
The data were reduced using the \textsc{ESORex} pipeline to extract, wavelength calibrate, and flat field each spectrum and form a data cube. The final cube was generated by aligning the individual observations and then median combining them; 3$\sigma$ clipping was applied to reject cosmic rays.
The seeing during these observations was insufficient to resolve the individual knots in system \#1 but the overall velocity structure is well sampled.

The lower panel of Fig.~\ref{fig:donut} shows the velocity maps extracted from fitting these spectra on the spaxel scale. The H$\alpha$ and [NII] line complex is clearly detected; we measure a velocity gradient of -50 to +90\,km s$^{-1}$ is recovered.

\section{Strong-Lensing Mass Model}
\label{sec:slmodel}
\begin{table}
\begin{center}
\caption{ Multiply imaged systems considered in this work, i.e. the (\emph{gold} + \emph{sylver}) set from \citet{mahler18b}.
The different star formation knots used as constraints for system \#1 and system \#2 are also highlighted on \emph{HST} colour composite stamps in Fig.~\ref{fig:sfknots}.
The flux magnification factors come from our best-fit mass model, with errors derived from MCMC sampling.
Redshifts with error bars correspond to values estimated by our model, errors come from MCMC sampling.
}
\begin{tabular}{lcccc}
\hline
\hline
ID & R.A. & Dec & $z$ & $\mu$\\
   & ($deg$) & ($deg$) & & \\
\hline
\hline
  1a.1 & 64.39615785 & -11.90676027 & 0.8718 & $5.4\pm1.8$\\
  1a.2 & 64.39431017 & -11.90713584 & 0.8718 & $3.9\pm1.5$\\
  1a.3 & 64.39034846 & -11.91086438 & 0.8718 & $2.5\pm1.0$\\
  1b.1 & 64.3960815 & -11.90725472 & 0.8718 & $7.1\pm2.1$\\
  1b.2 & 64.39472869 & -11.9075832 & 0.8718 & $4.0\pm1.5$\\
  1b.3 & 64.39029891 & -11.91127399 & 0.8718 & $2.3\pm0.9$\\
  1b.4 & 64.3949866 & -11.90892984 & 0.8718 & -- \\
  1c.1 & 64.39636429 & -11.90698317 & 0.8718 & $5.2\pm1.8$\\
  1c.2 & 64.39437083 & -11.9074093 & 0.8718 & $4.0\pm1.5$\\
  1c.3 & 64.39048794 & -11.911052 & 0.8718 & $2.5\pm0.9$\\
  1m.1 & 64.39616667 & -11.90692935 & 0.8718 & $5.7\pm1.9$\\
  1m.2 & 64.39443065 & -11.90727847 & 0.8718 & $4.9\pm1.7$\\
  1m.3 & 64.3903625 & -11.91101333 & 0.8718 & $2.4\pm0.9$\\
  2a.1 & 64.39909584 & -11.90636889 & 1.0460 & $2.6\pm1.0$\\
  2a.2 & 64.39556667 & -11.91118194 & 1.0460 & $2.5\pm1.0$\\
  2a.3 & 64.39137084 & -11.91207389 & 1.0460 & $3.1\pm1.2$\\
  2b.1 & 64.39899999 & -11.9066325 & 1.0460 & $2.7\pm1.1$\\
  2b.2 & 64.39582084 & -11.91122639 & 1.0460 & $2.5\pm1.0$\\
  2b.3 & 64.3912625 & -11.91232361 & 1.0460 & $2.9\pm1.2$\\
  2c.1 & 64.39900417 & -11.90685487 & 1.0460 & $2.8\pm1.1$\\
  2c.2 & 64.39595416 & -11.91129862 & 1.0460 & $2.5\pm1.0$\\
  2c.3 & 64.39129999 & -11.91249333 & 1.0460 & $2.8\pm1.1$\\
  3.1 & 64.39318 & -11.901537 & 1.0460 & $6.0\pm1.9$\\
  3.2 & 64.390026 & -11.903434 & 1.0460 & $10.4\pm2.5$\\
  3.3 & 64.388304 & -11.905013 & 1.0460 & $7.6\pm2.2$\\
  4.1 & 64.39952083 & -11.90747917 & $2.7\pm0.2$ & $6.5\pm2.0$\\
  4.2 & 64.39852916 & -11.90983889 & -- & $2.5\pm1.0$\\
  4.3 & 64.38609459 & -11.9153594 & -- & $2.1\pm0.8$\\
  5.1 & 64.379941 & -11.897906 & $2.0\pm0.2$ & $5.2\pm1.8$\\
  5.2 & 64.38237 & -11.896413 & -- & $7.6\pm2.2$\\
  5.3 & 64.388438 & -11.89163 & -- & $3.7\pm1.4$\\
  6.1 & 64.379991 & -11.897349 & $2.0\pm0.2$ & $6.5\pm2.0$\\
  6.2 & 64.381808 & -11.89639 & -- & $3.2\pm1.3$\\
  6.3 & 64.388558 & -11.89117 & -- & $3.4\pm1.3$\\
  8.1 & 64.388372 & -11.894492 & $2.3\pm0.4$ & $14.7\pm2.9$\\
  8.2 & 64.386885 & -11.895489 & -- & $18.8\pm3.2$\\
  9.1 & 64.382068 & -11.899994 & $7.6\pm0.5$ & -- \\
  9.2 & 64.382338 & -11.899779 & -- & $>20$ \\
  10.1 & 64.39839707 & -11.9071427 & $2.4\pm0.4$ & $8.4\pm2.3$\\
  10.2 & 64.39778511 & -11.90911404 & -- & $5.3\pm1.8$\\
  11.1 & 64.40154413 & -11.91891178 & $5.3\pm0.7$ & $>20$\\
  11.2 & 64.39970821 & -11.92009933 & -- & $>20$\\
  12.1 & 64.396902 & -11.897085 & $3.4\pm0.5$ & $2.4\pm0.9$\\
  12.2 & 64.38864 & -11.9013 & -- & $2.6\pm1.0$\\
  12.3 & 64.383172 & -11.906519 & -- & $2.9\pm1.1$\\
  13.1 & 64.397312 & -11.897068 & $3.6\pm0.5$ & $2.3\pm0.9$\\
  13.2 & 64.38842 & -11.901684 & -- & $2.6\pm1.0$\\
  13.3 & 64.38349928 & -11.9064465 & -- & $3.1\pm1.2$\\
  15.1 & 64.378193 & -11.89451 & $1.8\pm0.2$ & $5.4\pm1.8$\\
  15.2 & 64.38189 & -11.892331 & -- & $7.8\pm2.2$\\
  15.3 & 64.385361 & -11.890071 & -- & $5.6\pm1.9$\\
  16.1 & 64.385599 & -11.886984 & $3.2\pm0.8$ & $3.3\pm1.3$\\
  16.2 & 64.380143 & -11.888425 & -- & $1.5\pm0.5$\\
  16.3 & 64.376525 & -11.89254 & -- & $5.4\pm1.8$\\
  17.1 & 64.388212 & -11.895269 & $2.2\pm0.4$ & $>20$\\
  17.2 & 64.387833 & -11.895536 & -- & $>20$\\
  \hline
\hline
\end{tabular}
\label{tab:immul}
\end{center}
\end{table}

\subsection{Multiple images}
\label{sec:immul}
\cite{mahler18b} identified 57 multiple images (in 17 families) in the cluster core. They classified each image as \emph{gold}, \emph{silver} or \emph{bronze} following a similar process as for the \emph{Hubble Frontier Fields} \citep[HFF, ][]{lotz17}: \emph{(i) gold} images have spectroscopic redshifts, \emph{(ii) silver} images are securely identified through their geometry, colour, and morphology, and \emph{(iii) bronze} images are tentative identifications, some of which are based solely on predictions from the mass model.
In this analysis, we restrict ourselves to the (\emph{gold} + \emph{silver}) set, which consists of 15 systems comprising 41 individual images. We refer to this mulitple image set as \emph{silver} for the rest of the paper. IDs, coordinates, and redshifts for these multiple images are listed in Table~\ref{tab:immul}. Note that, for systems \#1 and \#2, images are decomposed into several star-formation knots as shown in Appendix~\ref{sec:sfknots} (Fig.~\ref{fig:sfknots}). As a result, we are using a total of 56 strong-lensing constraints.

\paragraph*{Redshift constraints --}
From the MUSE observations we obtain a spectroscopic redshift confirmation for 3 systems (systems \#1, \#2, and \#3). 
All four multiple images of system \#1, nicknamed \emph{The Doghnut}, are spectroscopically confirmed at $z=0.8718$. The spectra of the three most prominent images of system \#1 all show strong [OII], H$\beta$, H$\gamma$, H$\epsilon$ and H$\delta$ emission lines and MgII and Fe absorption lines in the UV. \emph{HST} and MUSE stamps are shown in the middle panel of Fig.~\ref{fig:donut}. The MUSE spectroscopic measurement is in good agreement with the \cite{mahler18b} LDSS3 spectroscopic redshift obtained at the Magellan Clay telescope (PI: Sharon) of $z=0.871$ for image 1.3 but with substantially better spectral resolution. The third row of Fig.~\ref{fig:donut} shows the MUSE velocity maps for all four images of system \#1 extracted by fitting the [OII] doublet directly, which can be compared to the high resolution SINFONI velocity maps calculated from fits to the [NII] and H$\alpha$ complex shown in the bottom row of Fig.~\ref{fig:donut}. The orientation of the observed rotation in the MUSE and SINFONI velocity maps is perpendicular to the critical line so the velocity maps appear to be very similar but one should note the low resolution MUSE velocity map of the third image (third stamp on third row of Fig.~\ref{fig:donut}) shows a different axis of rotation as expected from the lensing configuration.

System \#2 is composed of three multiple images, all spectroscopically confirmed at $z=1.046$ with strong [OII] and weak stellar absorption lines of the galaxy. All three images of system \#3 are also spectroscopically confirmed by MUSE at $z=1.046$ but due to the faint nature of this arc only [OII] is detected. As one can notice systems \#2 and \#3 are both at the same redshift. We discuss this in more details in Appendix~\ref{sec:galpair}. 
Finally we could not measure a secure redshift for system \#4, either $z=2.7$ or $z=3.1$ depending on the spectral feature associated with the deepest absorption line at 5720\AA. We discuss the case of system \#4 in more detial in Sect.~\ref{sec:sys4}.
Top panel of Fig.~\ref{fig:donut} shows a zoom into the core of MACS\,J0417, highlighting the geometrical configuration of system \#1 and system \#2 around the BCG.

\paragraph*{The fourth image of \emph{The Doghnut} --}
MUSE observations also reveal a fourth image for system \#1 which stands behind the brightest cluster galaxy (BCG), and thus is difficult to identify on the \emph{HST} imaging as shown in the second row of Fig.~\ref{fig:donut}. However, this image is partial, meaning only a part of \emph{The Doghnut} is being quadruply-imaged by the cluster. We include this fourth component of system \#1 in our mass model, and discuss its implications on the \emph{best-fit} mass model in Sect.~\ref{sec:compsl}. 

\begin{table*}
\begin{center}
\caption{PIEMD best-fit parameters for all three models that we run: the \emph{Fiducial Model}, the \emph{Variation 1 Model} and the \emph{Variation 2 Model}. $DM1$ refers to the main cluster halo, and $DM2$ and $DM3$ to the North and North West group-scale halos respectively. We refer to the BCG of the main cluster halo as $BCG1$, and to the massive ellipticals at the center of the North and North West group-scale halos as $BCG2$ and $BCG3$ respectively. For the scaling relation, the reference magnitude is $mag_{\rm F814W} =  19.45$.
$^{\rm a}$ Coordinates are given in arcseconds relative to the reference coordinate point: ($\alpha=64.394552$, $\delta=-11.908871$).
$^{\rm b}$ As explained in Sect.~\ref{sec:slres}, ellipticity ($e$) is defined to be $(a^2-b^2)/(a^2+b^2)$, where $a$ and $b$ are the semi-major and semi-minor axes of the ellipse.
$^{\rm c}$ The position angle $\theta$ is given in degrees and is defined as the direction of the semi-major axis of the iso-potential, counted counterclockwise from the horizontal axis (being the R.A. axis).}
\begin{tabular}{lrcccccccc}
\hline 
\hline
Model name & Component & $\Delta$x$^{\rm ~a}$ & $\Delta$y$^{\rm ~a}$ & $e^{\rm ~b}$ & $\theta^{\rm c}$& $\sigma$ & r$_{\rm cut}$ & r$_{\rm core}$\\ 
(Fit statistics) & -- & (\arcsec) & (\arcsec) &   & ($\deg$) & (km.s$^{-1}$) & (kpc) & (kpc)\\ 
\hline 
\hline 
\emph{Fiducial Model} & DM \#1 & 0.7$^{+0.4}_{-0.4}$ & 0.4$^{+0.8}_{-0.6}$ & 0.72$\pm$0.03 & 55.2$^{+0.5}_{-0.6}$ & 1033$^{+27}_{-22}$ & [1000.0] & 14.9$^{+0.7}_{-0.8}$\\ 
& DM \#2 & 44.4$^{+0.9}_{-2.0}$ & 71.8$^{+1.0}_{-1.6}$ & 0.67$\pm$0.05 & 44.2$^{+8.3}_{-5.5}$ & 499$^{+121}_{-28}$ & [1000.0] & 11.0$^{+6.4}_{-1.5}$\\ 
 & DM \#3 & 47.1$^{+2.6}_{-0.1}$ & 46.8$^{+0.8}_{-0.9}$ & 0.53$\pm$0.09 & 33.9$^{+14.4}_{-14.6}$ & 426$^{+33}_{-55}$ & [1000.0] & 12.5$^{+2.0}_{-3.8}$\\ 
 & BCG \#1 & [0.0] & [0.1] & [0.64] & [60.5] & 418$^{+19}_{-13}$ & 31.1$^{+14.5}_{-9.1}$ & [0.5]\\ 
 & BCG \#2& [47.8] & [69.6] & [0.35] & [74.1] & 311$^{+26}_{-39}$ & 102.3$^{+5.0}_{-20.5}$ & [0.3]\\ 
 & BCG \#3 & [46.9] & [48.4] & [0.16] & [50.6] & 269$^{+41}_{-15}$ & 66.8$^{+10.4}_{-13.2}$ & [0.0]\\ 
 & $L^{*}$ Galaxy & -- & -- & -- & -- & 209$^{+45}_{-27}$ & 15.5$^{+24.7}_{-3.3}$ & --\\ 
\hline

\emph{Variation 1 Model} & DM \#1 & 1.3$^{+0.3}_{-0.5}$ & 1.3$^{+0.5}_{-0.7}$ & 0.73$\pm$0.03 & 54.7$^{+0.7}_{-0.5}$ & 1001$^{+15}_{-19}$ & [1000.0] & 17.2$^{+0.4}_{-2.1}$\\ 
 & DM \#2 & 47.1$^{+0.9}_{-0.7}$ & 74.6$^{+2.0}_{-1.0}$ & 0.64$\pm$0.11 & 45.7$^{+4.9}_{-7.3}$ & 561$^{+58}_{-60}$ & [1000.0] & 16.5$^{+0.7}_{-1.8}$\\ 
 & DM \#3  & 46.2$^{+2.9}_{-1.6}$ & 44.8$^{+2.0}_{-1.7}$ & 0.42$\pm$0.08 & 34.5$^{+17.2}_{-11.8}$ & 439$^{+50}_{-41}$ & [1000.0] & 12.7$^{+1.7}_{-2.4}$\\
 & BCG \#1 & [0.0] & [0.1] & [0.64] & [60.5] & 473$^{+13}_{-37}$ & 67.6$^{+2.8}_{-16.1}$ & [0.5]\\ 
 & BCG \#2 & [47.8] & [69.6] & [0.35] & [74.1] & 305$^{+21}_{-50}$ & 64.1$^{+8.7}_{-27.8}$ & [0.3]\\ 
 & BCG \#3 & [46.9] & [48.4] & [0.16] & [50.6] & 247$^{+34}_{-12}$ & 17.2$^{+13.9}_{-9.9}$ & [0.0]\\ 
 & $L^{*}$ Galaxy & -- & -- & -- & -- & 209$^{+39}_{-34}$ & 12.3$^{+14.1}_{-12.6}$ & --\\ 
\hline 

\emph{Variation 2 Model} & DM \#1 & 0.1$^{+0.4}_{-0.5}$ & -0.2$^{+0.5}_{-0.5}$ & 0.69$\pm$0.03 & 54.2$^{+0.5}_{-0.4}$ & 981$^{+16}_{-28}$ & [1000.0] & 14.4$^{+1.0}_{-1.5}$\\ 
 & DM \#2 & 44.7$^{+2.2}_{-1.1}$ & 69.4$^{+1.9}_{-0.9}$ & 0.75$\pm$0.09 & 45.0$^{+4.0}_{-13.4}$ & 570$^{+51}_{-101}$ & [1000.0] & 17.3$^{+0.1}_{-1.7}$\\ 
 & DM \#3 & 46.3$^{+0.9}_{-1.2}$ & 53.5$^{+1.2}_{-2.4}$ & 0.74$\pm$0.13 & 53.7$^{+4.9}_{-7.0}$ & 397$^{+113}_{-33}$ & [1000.0] & 13.9$^{+4.3}_{-2.0}$\\ 
 & BCG \#1 & [0.0] & [0.1] & [0.64] & [60.5] & 417$^{+21}_{-16}$ & 85.2$^{+2.9}_{-13.5}$ & [0.5]\\ 
 & BCG \#2 & [47.8] & [69.6] & [0.35] & [74.1] & 375$^{+32}_{-21}$ & 98.2$^{+10.4}_{-31.7}$ & [0.3]\\ 
 & BCG \#3 & [46.9] & [48.4] & [0.16] & [50.6] & 276$^{+16}_{-22}$ & 25.3$^{+18.5}_{-12.4}$ & [0.0]\\ 
 & $L^{*}$ Galaxy & -- & -- & -- & -- & 88$^{+28}_{-27}$ & 89.2$^{+9.9}_{-56.7}$ & --\\ 
\hline 
\hline
\end{tabular}
\label{tab:slres}
\end{center}
\end{table*}
\begin{table*}
\begin{center}
\caption{Figures of merit for each model considered in Sect.~\ref{sec:compsl}. Columns show the Bayesian evidence ($\log{E}$) and likelihood ($\log{\mathcal{L}}$), the rms deviation of predicted multiple-image positions from their observed positions in the image plane, $rms$, the reduced $\chi^2$, $\chi_{red}^{2}$, the Bayesian Information Criterion, BIC, the Akaike Information Criterion, AIC, and the corrected AIC, AICc.
We also quote the improvement on several parameters compare to the fiducial model: on the the BIC, $\delta_{\rm BIC}$, and the AICc, $\delta_{\rm AICc}$. A value of $\delta_{\rm BIC}$ and $\delta_{\rm AICc}$ greater than 10 reflects a significant improvement/degradation of the model compare to the fiducial one.
}
\begin{tabular}{ccccccccccc}
\hline
\hline
Model & $\log{E}$ & $\log{\mathcal{L}}$ & $rms$ & $\chi^2_{\rm red}$ & BIC & AIC & AICc & $\delta_{\rm BIC}$ & $\delta_{\rm AICc}$\\
\hline
\hline
\emph{Fiducial Model} & -104 & -41 & 0.38\arcsec & 0.9 & 262 & 158 & 198 & -- & -- \\
\hline
\emph{Variation 1 Model} & -110 & -38 & 0.34\arcsec & 0.8 & 254 & 151 & 192 & 8 & 6 \\
\emph{Variation 2 Model} & -117 & -48 & 0.45\arcsec & 1.3 & 271 & 171 & 209 & -9 & -11 \\
\hline
\hline
\end{tabular}
\label{tab:figmer}
\end{center}
\end{table*}

\subsection{Methodology}
\label{sec:methodo}
To model the mass distribution of MACS\,J0417 we use the \textsc{lenstool} software \citep{jullo07}. Our method closely follows the method used in previous works \citep{jauzac14,jauzac15b,jauzac16a} so we here only give a brief summary of the different steps in the build-up of the mass model. We refer the reader to \cite{kneib96}, \cite{smith05}, and \cite{richard10,richard14} for more details.

Our mass model combines large-scale dark matter halos to model the cluster components, and small-scale dark matter halos to model the cluster galaxies, typically large ellipticals (like the BCG) and galaxies in the proximity of multiple-images. All mass components are modelled as Pseudo Isothermal Elliptical Mass Distribution \citep[PIEMD, ][]{limousin05,eliasdottir07}, parametrized by a position ($x$,$y$), an ellipticity ($e$), a position angle ($\theta$), a velocity dispersion ($\sigma$), a core radius ($r_{\rm core}$), and a cut radius ($r_{cut}$). 
For the PIEMDs used to model small-scale mass perturbers, cluster galaxies, we fix the parameters ($x$,$y$), $e$, and $\theta$, at the values measured from their light distribution \citep[][]{kneib96,limousin07b,richard14} and assume a Faber-Jackson empirical scaling relation \citep{FJ76} to relate their velocity dispersion and cut radius to their observed luminosity \citep{jauzac16a}. We optimize the velocity dispersion and the cut radius for an $L^{\ast}$ galaxy only with $mag_{0}=19.45$ in the ACS/F814W passband. The velocity dispersion is allowed to vary between 20 and 220\,km.s$^{-1}$, and the cut radius between 1 and 100\,kpc. The r$_{\rm cut}$ upper limit is there to account for tidal stripping of galactic dark matter halos \citep{limousin07a,limousin09a,natarajan09,wetzel10}.
With such a parametric approach, dark matter halos are not allowed to contain zero mass, and hence we measure the goodness of our model to meet the observational constraints with the $\chi^2$ and rms statistics. The rms is measured as the difference between the observed position of the multiple images and the predicted position from the model. In principle, a low rms would indicate a better model (see Sect.~\ref{sec:sys4}).

\begin{center}
\begin{figure*}
\includegraphics[width=0.495\textwidth,angle=0.0]{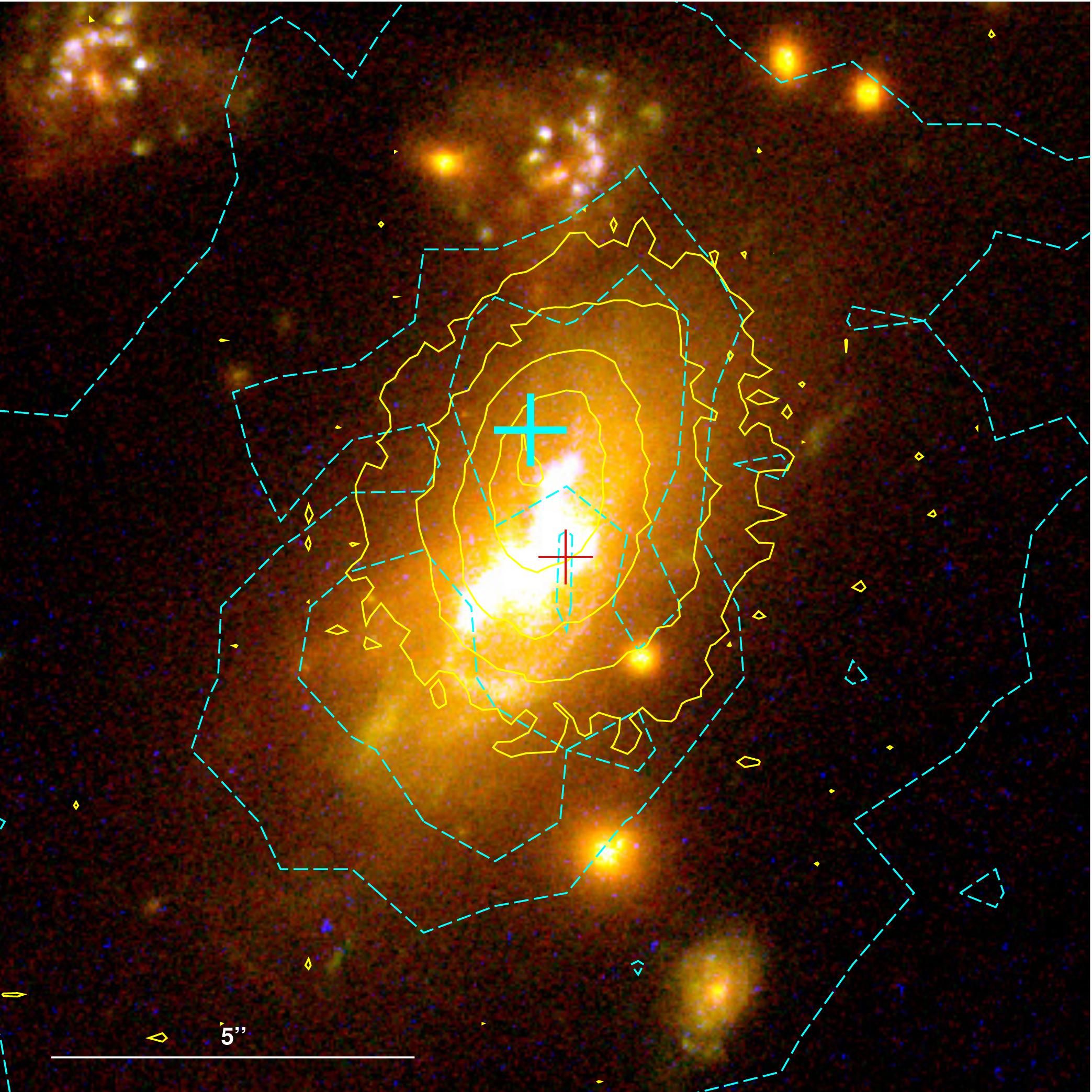}\ 
\includegraphics[width=0.495\textwidth,angle=0.0]{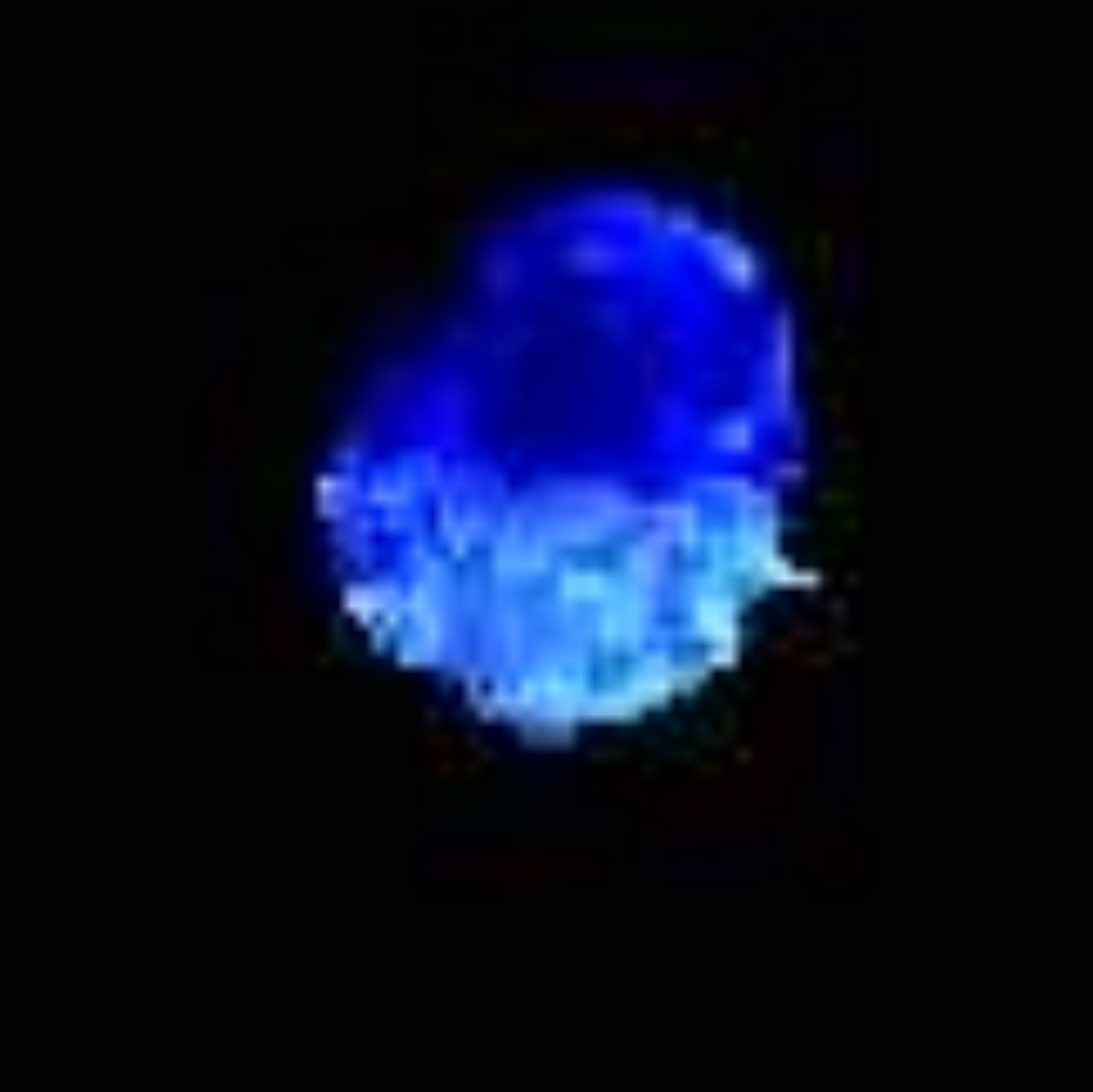}\ 
\\
\caption{\textit{Left panel: }\emph{HST} colour composite stamp centered on the BCG. Cyan contours show the X-ray gas distribution from \emph{Chandra}. Yellow contours show the [OII] emission of the BCG. The cyan and red crosses highlight the gas peak from the X-ray and the dark matter peak from the lensing analysis respectively. One will note the [OII] emission peak is consistent with the gas peak from the X-ray. The stamp is 15\arcsec$\times$15\arcsec.
\textit{Right panel: }MUSE low-resolution [OII] velocity map of the BCG. The stamp size is the same as the left panel and covers a velocity range of -450 to +450\,km.s$^{-1}$. There is a significant blueshift to the optical line emission with gas velocities ranging from -50 to -450\,km.s$^{-1}$, and a prominent velocity gradient across the system along the major axis of the BCG.
}
\label{fig:bcg}
\end{figure*}
\end{center}

\subsection{Results}
\label{sec:slres}
Our final mass model includes three cluster-scale halos. MACS\,J0417 is a dynamically active cluster, which can be separated into three main components: the main halo located around the BCG ($\alpha=64.394552$, $\delta=-11.908871$), and two other group-scale halos located North ($\alpha=64.380985$, $\delta=-11.889541$, and North West ($\alpha=64.381231$, $\delta=-11.895417$) of the main cluster. Both group-scale halos have an over-density of cluster members located in their surroundings with a bright elliptical galaxy at their centre. 
For all three large-scale halos, we allow their positions to vary within 10\arcsec of their associated light peak, and their core radius between 1 and 20\arcsec. The velocity dispersion of the main halo can vary between 500 and 2\,000\,km.s$^{-1}$, and the velocity dispersion of the group-scale halos between 200 and 1\,000\,km.s$^{-1}$. The ellipticity, defined as $e=(a^2-b^2)/(a^2+b^2)$, and position angle of all three are also free parameters during the optimization process. The cut radius for all three is fixed to 1\,000\,kpc as strong-lensing constraints alone cannot probe the outer regions of the dark matter potential.
Additionally, we include 177 mass perturbations induced by cluster galaxies by assigning to them a galaxy-scale halo following the method presented in Sect.~\ref{sec:methodo} \citep{richard14}. Finally, we add three galaxy-scale halos to model the BCG and the two ellipticals at the centre of the main, North and North-West large-scale halos respectively for which we optimize both their cut radius and velocity dispersion.
Using the 56 multiple images presented in Sect~\ref{sec:immul} and listed in Table~\ref{tab:immul}, we optimize the above-mentioned free parameters of our mass model using \textsc{lenstool}.
The redshift parameter of each multiply-imaged system without a spectroscopic measurement was set as a free parameter, and allowed to vary between 0.6 and 9.

The best-fit mass model optimized in the image plane gives us a rms of 0.38\arcsec and a reduced $\chi^2$ of 0.9. The best-fit redshifts for the multiple images and their estimated magnifications are given in Table~\ref{tab:immul}. 
Our rms of 0.38\arcsec is similar to that obtained by \cite{mahler18b}, 0.37\arcsec. This excellent agreement in terms of rms is encouraging as both teams have modeled the mass differently, i.e. using a different prior mass distribution with two more group-scale halo components to model the North and North West clumps, and adding the fourth partial image of \emph{The Doughnut} on our side. The comparison between these two models is discussed in Sect.~\ref{sec:comp_muserel}.

\begin{figure*}
\begin{center}
\includegraphics[width=\textwidth]{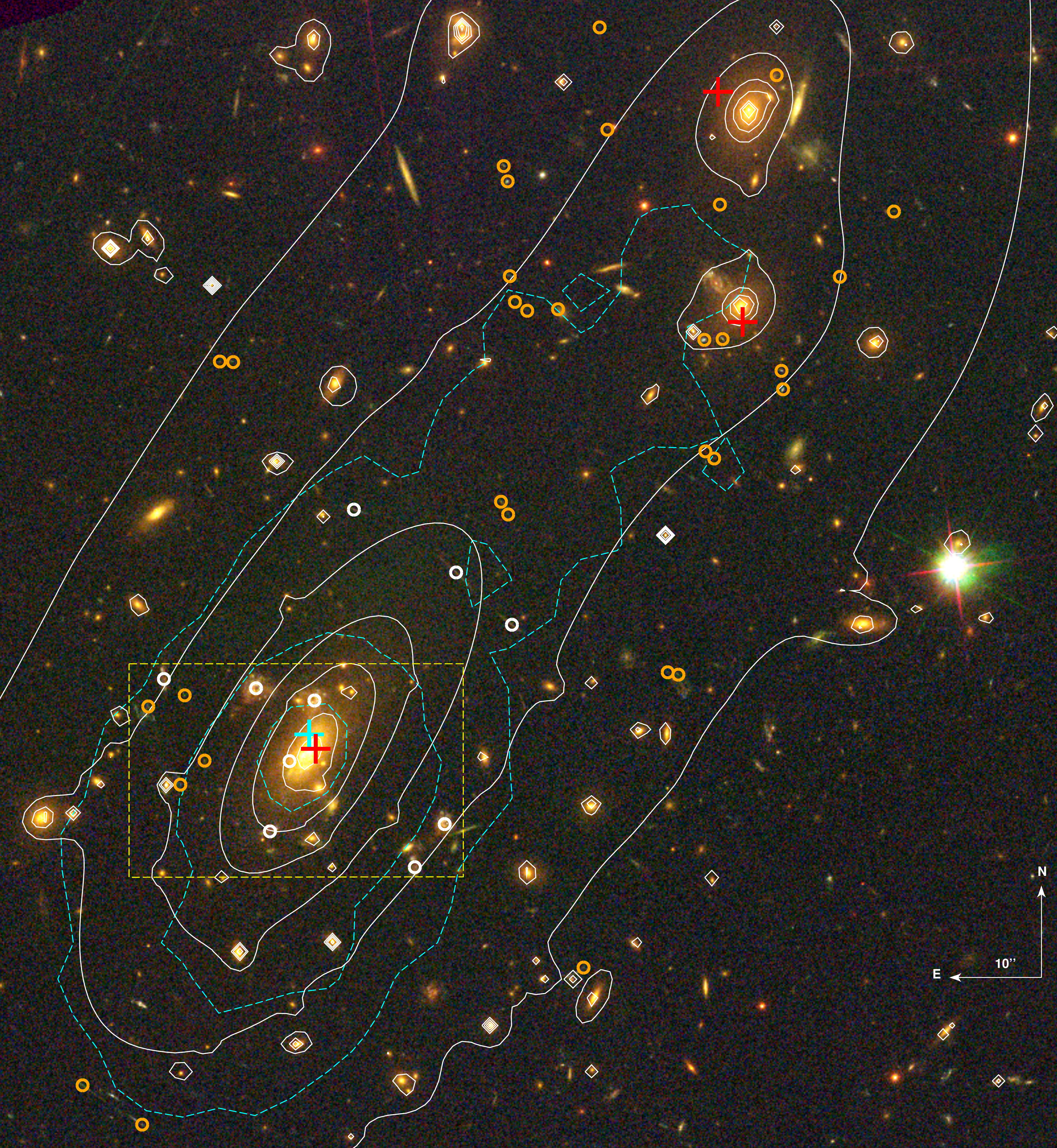}\\
\caption{Colour composite image created from HST/ACS and WFC3/UVIS images in the F814W, F606W, and F435W passbands. Orange circles mark the positions of multiple images as listed in Table~\ref{tab:immul}. White circles highlight multiple images spectroscopically confirmed in VLT/MUSE observations. Red crosses show the positions of the dark-matter clumps of the \emph{Fiducial Model}, and the cyan cross is placed at the position of the X-ray peak. White and the dashed cyan contours show the distribution of mass reconstructed from our lensing mass model and gas from \emph{Chandra} observations, respectively. The yellow box highlights the region visible in the top panel of Fig.~\ref{fig:donut}. We find the main dark-matter halo to be well aligned with the light peak of the BCG, but offset with respect to both the peak of the X-ray surface brightness and the peak of the optical line emission from the BCG.}
\label{fig:acs}
\end{center}
\end{figure*}

The best-fit parameters of our mass model are given in Table~\ref{tab:slres} under \emph{Fiducial Model}. One will note that while all three dark matter large-scale halos are allowed to move around their light peak, both the main and the North West halos are well-aligned with their respective BCG as can be seen in Fig.~\ref{fig:acs} (dark matter peaks are highlighted with red crosses). However such good alignement between dark matter and light is not observed for the North halo. This could be due to several factors, such as the dynamical status of the cluster for example which is further discussed in Sect.~\ref{sec:compsl}. However even if we include the North and the North West components in our model, none of the multiply-imaged systems in their vicinities are spectroscopically confirmed. This implies the model in this region is degenerated and therefore any conclusion should be discussed with care concerning the dark matter centering of the North and North West clumps.

In order to provide a two-dimensional (cylindrical) mass of MACS\,J0417, we integrate the mass map within annuli centered on the main BCG. We measure a mass within 200\,kpc of $M({\rm R<200\,kpc})=1.77\pm0.03\,10^{14}\,\msun$. The mass distribution obtained with this model is shown in Fig.~\ref{fig:acs} as white contours.

\section{Discussion}
\label{sec:discussion}

\subsection{The Brightest Cluster Galaxy}
\label{sec:bcg}
The most striking feature in MACS\,J0417 from the MUSE data is the disturbed BCG in the core of this cluster. It shows strong line emission (yellow contours in the left panel of Fig.~\ref{fig:bcg}) and significant H$\delta$ absorption consistent with the active star-formation first noted by \cite{green16}.
We see a significant spatial offset in the optical line emission from the stellar component of the BCG, $1.7\pm 0.5$\arcsec or 9.8$\pm 2.9$~kpc, that is consistent with the peak in the X-ray emission observed by {\it Chandra}. There is a significant blueshift to the optical line emission with gas velocities ranging from -50 to -450~km\,s$^{-1}$  and a prominent velocity gradient across the system along the major axis of the BCG that can be seen in the right panel of Fig.~\ref{fig:bcg}.
Similar spatial and dynamical offsets in optical line emission have been seen in a number of other clusters \citep{hamer12} that exhibit `sloshing' activity, and the one observed here in MACS\,J0417 is the most distant example of this phenomenon. The direction of the `sloshing' of the intra-cluster gas is consistent with the axis defined by the three mass components found in our lensing analysis and the more extended morphology of the cluster in the X-ray. Figure~\ref{fig:acs} shows the overall X-ray emission of the cluster as cyan contours, with the X-ray peak highlighted with a cyan cross. The dark matter peaks are shown as red crosses.

The fact that this offset gas cooling from the intra-cluster medium is coincident with the ionised cool gas points to direct deposition of cold molecular gas from the intra-cluster medium that is unrelated to the stellar population of the BCG. Therefore, in the cases where the X-ray peak is coincident with the BCG it is the cooling from the intra-cluster medium, and not stellar mass loss, that dictates the cold gas reservoir built up in the cluster core.

\subsection{Dynamical History of MACS\,J0417}
\label{sec:compsl}
We aim at measuring precisely the dark matter peak location in the cluster core, compare it with its light and gas counterparts, confirm the dynamical history of the cluster core, and possibly put constraints on dark matter particle's nature.
We first estimate the improvement of our mass model when including the fourth partial multiple image of \emph{The Doghnut}, and what it suggests about the dark matter centering in the main cluster halo. 
Adding to our \emph{Fiducial Model}, we run a second mass model which does not include this partial fourth image of system \#1, \emph{Variation 1 Model}, but remains identical in terms of prior mass distribution: three large-scale dark matter halos, all three respective BCGs as well as 177 cluster galaxies. The best-fit parameters are given in Table~\ref{tab:slres}. 

Following the comparison method used by \cite{acebron17}, \cite{lagattuta17}, \cite{mahler18a}, and \cite{jauzac18a}, we measure three different Bayesian estimates to compare our two models, adding to the standard rms and reduced $\chi^2$: the Bayesian Information Criterion (BIC), the Akaike Information Criterion (AIC), and the Akaike Information Criterion corrected (AICc).
The BIC is given by:
\begin{equation}
{\rm BIC}=-2\log(\mathcal{L})+k\log(N)\ ,
\end{equation}
where $N$ is the number of constraints and $k$ is the number of free parameters.

The Akaike Information Criterion can be calculated as: 
\begin{equation}
{\rm AIC}= - 2\log(\mathcal{L}) + 2\ k\ ,
\end{equation}
which is a more robust estimate of overfitting.

And finally the Akaike Information Criterion corrected is the AIC corrected for a finite number of free parameters:
\begin{equation}
{\rm AICc}={\rm AIC} + \frac{2\ k\ (k+1)}{(N - k - 1)}\ ,
\end{equation}

For the BIC, the AIC, and the AICc, a penalty term for models with too many free parameters that overfit noise rather than capture additional information is included, which is larger with BIC and AICc than with AIC.
Note that BIC, AIC and AICc were all developed for estimating the goodness of fits to models with linear parameters. As they provide us with an estimate, strong gravitational lensing is highly non-linear, thus these values should be interpreted with caution.
For all these figures of merit, including the $\chi^2$ and the rms, lower values should be preferred. An improvement of less than 10 on the AICc should not be considered as statistically significant \citep{jauzac18a}. They are given for each model in Table~\ref{tab:figmer}.
Improvement on all figures of merit between the \emph{Fiducial Model} and \emph{Variation 1 Model} are observed, however the improvement on the most reliable Bayesian estimator, the AICc, is less than 10, and is therefore considered as not statistically significant. 
	However, the inclusion of this fourth image has a noticeable impact on the position on the dark matter peak.
	
The central region of the cluster core, where the BCG resides, is dominated by the stellar distribution, therefore constraining the exact location of the dark matter peak is extremely difficult and the error bars are usually large (can reach up to a few arcseconds) if we cannot add strong-lensing constraints in this region of the cluster or measure the stellar velocity dispersion of the BCG with high resolution as was done by \cite{newman11,newman13a,newman13b} and \cite{monna15,monna17}.
The geometry of system \#1 is relatively common when looking at massive lenses, i.e. with a fourth radial image predicted in the central region \citep[e.g. ][]{sand04,sand08}. However due to the BCG being located in this region, it is difficult with usual optical data (\emph{HST} in most cases) to identify this counterpart and locate it exactly, even with BCG subtraction methods as those usually leave residuals. Therefore, the centre of the dark matter halo is subject to large uncertainties.  

In the case of MACS\,J0417, the MUSE data offer an opportunity to better constrain our mass model and precisely locate the main dark matter clump (DM \#1 in Table~\ref{tab:slres}) due to the identification of the fourth radial image of system \#1. Our \emph{Fiducial Model} predicts an alignment of $\pm$0.5\arcsec between the light peak (traced by the BCG) and the dark matter peak as shown in Fig.~\ref{fig:acs} (dark matter peaks are shown with red crosses). 
Such alignment is in good agreement with a cold dark matter \citep[CDM;][]{peebles84,blumenthal84} scenario where the dark matter potential is predicted to be cuspy, and light and dark matter are expected to `stick' together. 
In a scenario where self-interacting dark matter were to be involved \citep[SIDM;][]{spergel00}, the dark matter potential is predicted to have a core, allowing the BCG to `slosh' around, and therefore dark matter and  light are not expected to `stick' together anymore. \cite{kim17} demonstrated with SIDM simulations that offsets between light and dark matter peaks could reach up to a few hundred kpc. This was then observationally confirmed by \cite{harvey17}.
Therefore, measuring precisely the center of the dark matter potential and its offset with the light peak is one of the powerful tests that can be conducted in cluster cores to study the nature of dark matter \citep[for more tests and investigations see ][]{robertson18}. Our model without the fourth image (\emph{Variation 1 Model}) predicts an offset between light and dark matter in the main potential (DM \#1) of $1.3\pm0.5$\arcsec, equivalent to an offset of $7.4\pm2.9$\,kpc, which could be interpreted as being in favour of a SIDM scenario \citep{massey15,massey18}. On the other side, our \emph{Fiducial Model} predicts an offset of $0.5\pm0.5$\arcsec, consistent with a perfect alignment, and thus in favour of a CDM scenario.

In order to push our investigation further, we look at all three main components of the cluster: dark matter, light and gas. The X-ray analysis with \emph{Chandra} by \cite{ME12} suggests that MACS\,J0417 has an ongoing merger event, its gas peak being misaligned with the light \citep{ME12}, with a gas plume in the direction of the NW halo as shown in Fig.~\ref{fig:acs} with the cyan contours, tracing a former interaction between the main and the North-West halos.
The different components of MACS\,J0417 exhibit a familiar scenario observed in several other merging clusters such as \emph{the Bullet}, \emph{the baby-bullet} or Abell\,2744 and MACS\,J0717 \citep{clowe04,bradac06,bradac08b,owers11,jauzac16b,ma09,jauzac18b}: dark matter and stars being aligned with the gas lagging behind. Such a situation represents another way to put constraints on dark matter's nature.
Merging clusters and in-falling substructures have often been used as ways to probe the nature of dark matter through its self-interaction cross-section \citep{kahlhoefer14,harvey13}. As these in-falling halos pass through the dense environment of another galaxy cluster the three components (gas, dark matter and galaxies) experience differential forces causing the dynamical behaviour of each component to change. The gas for example feels a large drag force as it interacts with the gaseous intra-cluster medium, the galaxies feel only the dynamical gravitational friction of the lumpy medium and the dark matter will behave according to its nature. Should dark matter have a significant self-interaction cross-section the halo will drag similarly to that of the gas, temporarily separating from the galaxies \citep{harvey14,robertson17}. 
Our analysis of MACS\,J0417 shows an offset between light and dark matter of the main cluster halo (DM\#1) of $0.5\pm0.5\arcsec$ and a gas offset of $1.7\pm0.5\arcsec$. Adopting the method used in \cite{harvey15} and calibrated to simulations \citep[][Harvey et al.\ \emph{in prep.}]{robertson18} we can interpret this offset as caused by the particle properties of dark matter due to the former interaction between the main halo and the North-West halo of MACS\,J0417.
 After mass matching the simulations we find that the observed separation implies a cross-section upper-limit $\sigma_{\rm DM}/m < 3$\,cm$^2$.g$^{-1}$ consistent with other observations \citep[e.g.][]{bradac08b,dawson12}.

However, if we consider the significant offset in the optical emission line associated with the star-forming BCG in this cluster, and discussed in Sect.~\ref{sec:bcg}, we conclude that the observed offset between the gas and light/dark matter peaks in the main halo of MACS\,J0417 can be explained without including another possible dark matter candidate such as SIDM, which is in good agreement with the constraints we put on its self-interaction cross-section.

\begin{center}
\begin{figure*}
\includegraphics[width=0.495\textwidth,angle=0.0]{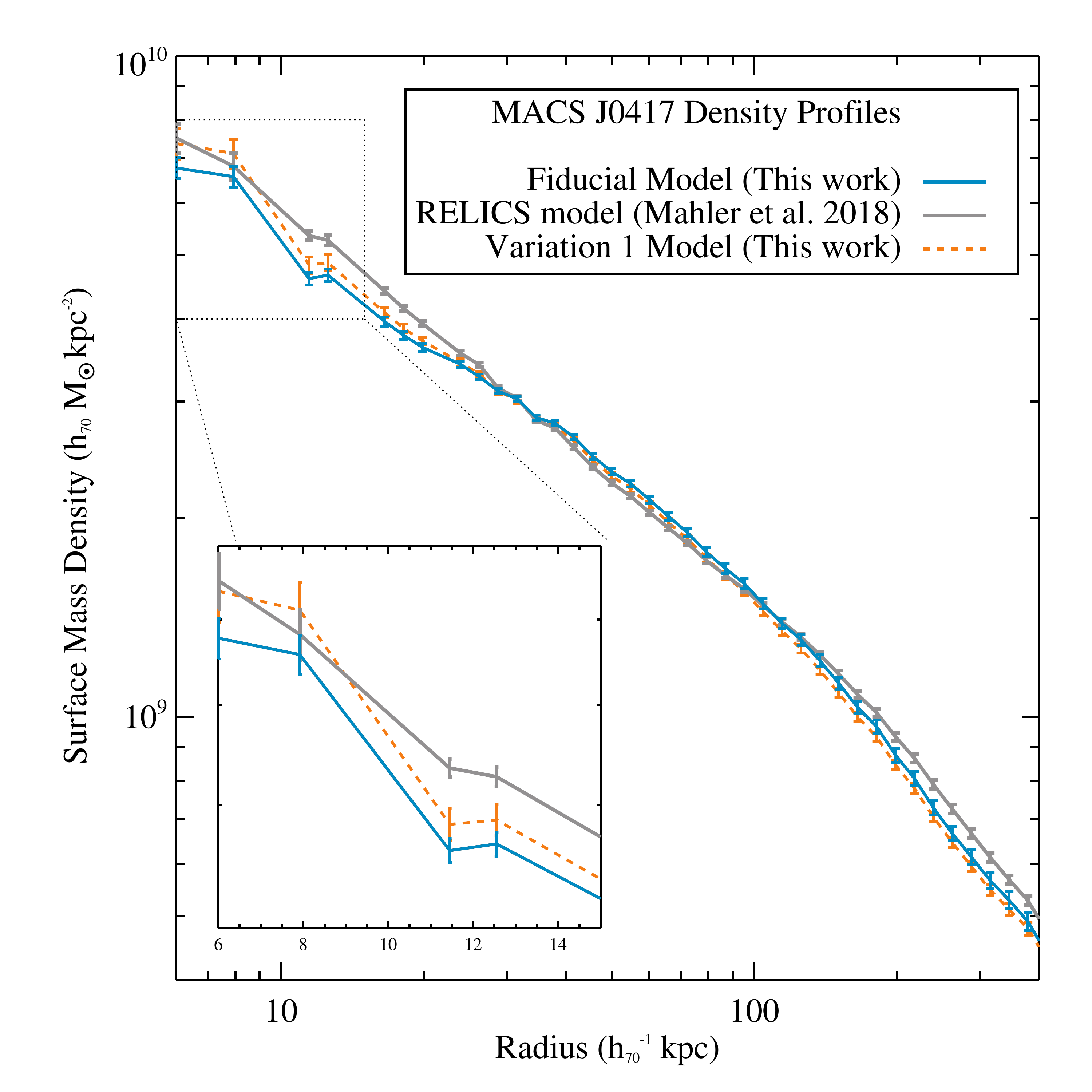}\ 
\includegraphics[width=0.495\textwidth,angle=0.0]{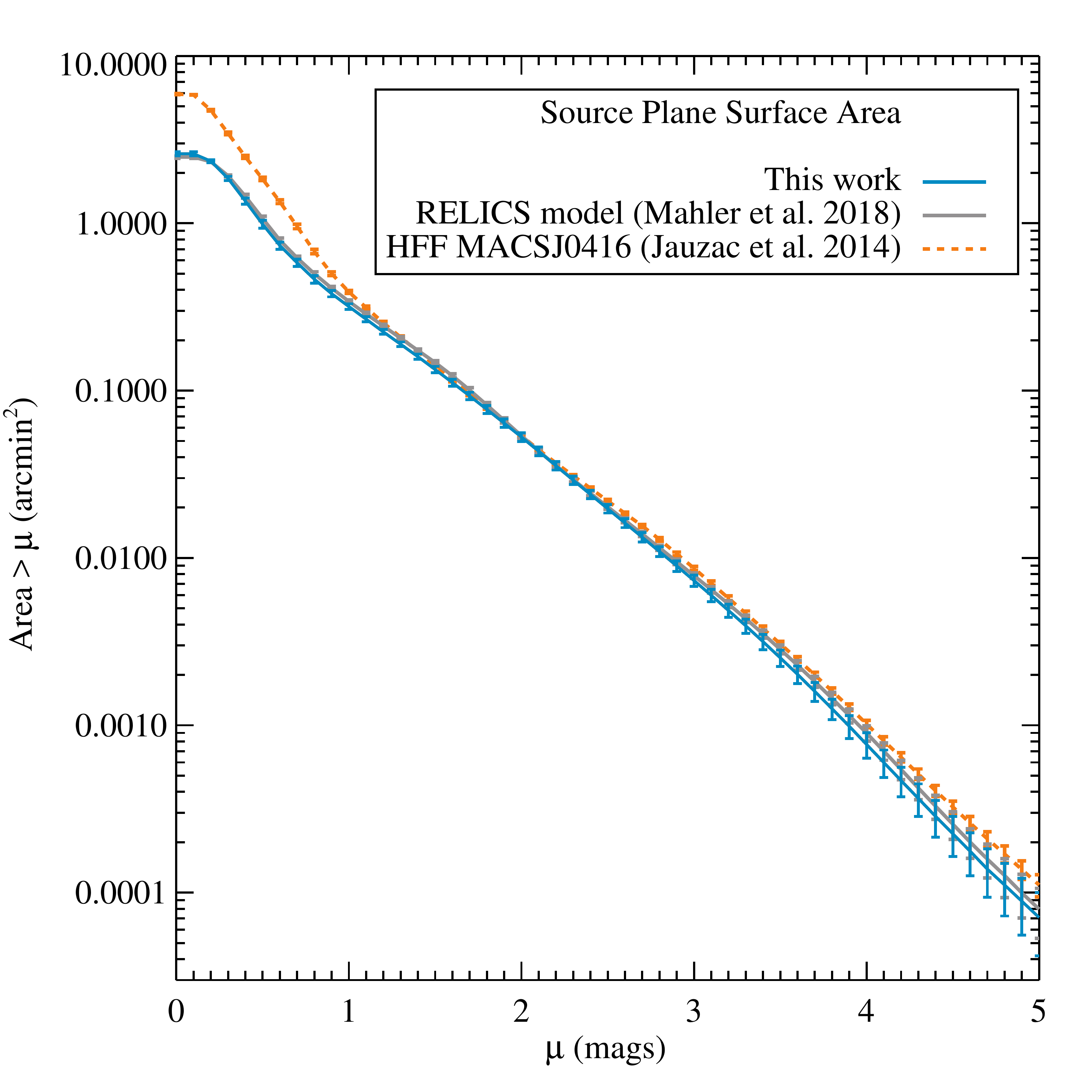}\ 
\\
\caption{\textit{Left panel: } 2D-density profiles obtained from our \emph{Fiducial Model} in cyan, from the RELICS analysis from \citet{mahler18b} in grey, and from our \emph{Variation 1 Model} in orange. One can see a really good agreement between all three models. The slight differences in the core is attributed to the presence of the fourth partial image of system \#1 on our side, better constraining the inner most $\sim$10\,kpc of the dark matter density profile. The inset plot zooms onto the 6-15\,kpc region of the density profiles. \textit{Right panel: }Surface area in the source plane covered by the overlap between ACS and WFC3 at a magnification above a given threshold $\mu$ for the present model in cyan, and the RELICS model presented in our companion paper by \citet{mahler18b} in grey. When put in perspective of MACS\,J0416 (orange dotted line), one of the HFF clusters, one can see that MACS\,J0417 has a high lensing power.
}
\label{fig:comp_muse-relics}
\end{figure*}
\end{center}

\subsection{The impact of the redshift of system \#4}
\label{sec:sys4}
Our \emph{Fiducial Model} does not include the redshift of system \#4 as its measurement from the MUSE data cube is uncertain as discussed in Sect.~\ref{sec:immul}. As explained in Sect.~\ref{sec:methodo}, multiply-imaged systems without a spectroscopic confirmation have their redshift being optimized. The best-fit redshifts are given in Table~\ref{tab:immul}. We do not use any photometric redshift prior, and for all systems without a spectroscopic redshift we let the redshift vary between 1 and 9 during the \textsc{lenstool} optimization. In the case of system \#4, one can see that our \emph{Fiducial Model} favors the 2.7 spectroscopic measurement, with a best-fit redshift of $2.7\pm0.2$ (also in agreement with \emph{Variation 1 Model} which gives an optimized redshift of $2.8\pm0.2$).

As a matter of check, we run another model which now includes a redshift of 2.7 for system \#4. The best-fit parameters of the model are given in Table~\ref{tab:slres}, and its figures of merit in Table~\ref{tab:figmer} under \emph{Variation 2 model}.
This model is identical to our \emph{Fiducial Model}, except we have now fixed the redshift of system \#4 to 2.7. The model is actually degraded compared to our \emph{Fiducial Model}: an increase of the rms of 0.08\arcsec, and an increase of the AICc of 11. The value of the $\delta_{\rm AICc}$ between our \emph{Fiducial Model} and the \emph{Variation 2 Model} is greater than 10 and is thus considered here as significant.
Adding the redshift of system \#4 means we remove a free parameter from the model, therefore lower its flexibility.

\cite{johnson16} studied the impact of the redshift information for multiple images on the resulting mass model. They showed that less flexible models (such as \emph{Variation 2 Model} in this work) due to more constraints (the redshift of system \#4 in our case) produce higher rms, and higher figures of merit in general as shown in Table~\ref{tab:figmer}, which indicates a degraded model fit compare to the \emph{Fiducial Model}. However, the investigation from \cite{johnson16} shows that such models are usually better at predicting the locations of images across the whole image plane. The comparison between our \emph{Fiducial Model} and \emph{Variation 2 Model} agrees with the conclusions from \cite{johnson16} and highlights the fact that standard/linear bayesian estimators of fit goodness should be taken with caution when it comes to comparing strong-lensing mass models, in particular if they use different number of constraints..

\subsection{Comparison with the \citet{mahler18b} mass model}
\label{sec:comp_muserel}
We compare our aperture mass measured within 200\,kpc to the one obtained by the RELICS team and presented in \cite{mahler18b}. 
In this Section, we refer to the \emph{Fiducial Model} as our model, and for the \cite{mahler18b} model we refer to their \emph{Silver Model}.
They measure a two dimensional integrated mass of $M({\rm R<200\,kpc})=1.78\pm0.02\times10^{14}\,\msun$ which is in excellent agreement with our measurement of $M({\rm R<200\,kpc})=1.77\pm0.03\times10^{14}\,\msun$.

The left panel of Fig.~\ref{fig:comp_muse-relics} shows the two dimensional density profiles obtained by our \emph{Fiducial Model} in cyan, and by the \cite{mahler18b} model in grey. One can see that the two profiles exhibit a similar trend, with our model predicting a lower density in the core. The only difference between the two models in this region of the cluster is the inclusion of the fourth image of system \#1, which is not included in the RELICS model. While the \cite{mahler18b} model predicts a fourth image for this system, its position cannot be unambiguously identified in the \emph{HST} data alone.The fourth multiple image of \emph{The Doghnut} is a radial pair merging on the critical curve, therefore it brings strong constraints on the inner density profile of the cluster potential ($<10$\,kpc) as discussed in Sect.~\ref{sec:compsl}. In Fig.~\ref{fig:comp_muse-relics}, we also plot the density profile we obtain with our \emph{Variation 1 Model} in orange, which is identical to the \emph{Fiducial Model} apart from the inclusion of the fourth image of system \#1. One can see that the inner ($R>10$\,kpc) density profile in this case is similar to the one obtained by \cite{mahler18b}. We thus conclude that this additional constraint is responsible for the difference in shape of the inner density profile of MACS\,J0417.

The right panel of Fig.~\ref{fig:comp_muse-relics} shows the surface area in the source plane, $\sigma_{\mu}$, above a given magnification factor, $\mu$, for our \emph{Fiducial Model} (cyan) and the RELICS model (grey) presented in our companion paper \citep{mahler18b}. This metric is a good estimator of the efficiency of the lensing configuration to magnify high-redshift galaxies as suggested initially by \cite{wong12}, as $\sigma_{\mu}$ is directly proportional to the unlensed comoving volume covered at this magnification. We compute $\sigma_{\mu}$ for a source at $z=9$. It is calculated from the multiple image with the highest magnification at each source position.
\cite{mahler18b} centered their analysis on a comparison of photometric and model-fitted redshifts as well as on the lensed high-redshift candidates for future follow-up with the \emph{James Web Space Telescope} (JWST). We use this Section to demonstrate that even if our models are slightly different, the outputs are in extremely good agreement. Following our work on the \emph{Hubble Frontier Fields}, we estimate the area above $\mu=3$ as our metric. We measure $\sigma(\mu>3)=0.22$\,arcmin$^{2}$ and $\sigma(\mu>3)=0.23$\,arcmin$^2$ for our \emph{Fiducial Model} and the RELICS model respectively.
These two values are in excellent agreement. Our measurement of $\sigma(\mu>3)$ can be put in perspective with the values obtained for some of the HFF clusters, i.e. 0.44\,arcmin$^2$ for Abell\,2744 \citep{jauzac15b}, 0.26\,arcmin$^2$ for MACS\,J0416.1-2403 \citep{jauzac14}. We show in the right panel of Fig.~\ref{fig:comp_muse-relics} the same relation but for MACS\,J0416 in orange. This strengthens the fact that MACS\,J0417 has a relatively high lensing power, and thus should be an interesting cluster to probe the high-redshift universe as stressed by \cite{mahler18b}.

\section{Conclusion}
\label{sec:conclusion}
We present MUSE observations of the massive galaxy cluster MACS\,J0417.5-1154 at $z=0.441$,  discovered in the MAssive Cluster Survey \citep[MACS,][]{ebeling01,ME12}.
The high mass and disturbed morphology of MACS\,J0417 make it a powerful gravitational lens. VLT/MUSE observations provide spectroscopic confirmation of three of the four high-confidence  multiple-image families in the MUSE field of view: system \#1 at $z=0.8718$, systems \#2 and \#3 at $z=1.046$, and system \#4 at either $z=2.7$ or $z=3.1$, where the MUSE spectral range contains few strong lines. System \#1 is of particular interest, as the MUSE observations provide us with the detection of a fourth radial image located behind the BCG, which would have been difficult to locate from imaging data alone (even at \emph{HST} resolution), as the BCG's high star-formation rate contaminates the bluer bands where the fourth image of system \#1 is visible.

Based on the multiple images identified in our companion paper \citep{mahler18b}, we build a strong-lensing mass model including the MUSE redshifts and the fourth image of system \#1, referred to as the \emph{Fiducial Model}. Our best-fit mass model recovers the multiple-image positions with an rms of 0.38\arcsec and yields a two-dimensional enclosed mass of $M({\rm R<200\,kpc})=1.77\pm0.03\times10^{14}\,\msun$. 
We then run a test model, called the \emph{Variation 1 Model}, which does not include the central image of system \#1. Unlike the \emph{Fiducial Model}, the \emph{Variation 1 Model} does not predict almost perfect alignment between the light peak of the BCG and the dark-matter peak. This difference highlights the impact and thus the importance of strong-lensing radial constraints for precise measurements of the location of the dark-matter peak in cluster cores. The good alignment of $(0.5\pm0.5)$\arcsec\ found by the \emph{Fiducial Model} is consistent with a cold dark-matter scenario. 
The identification of this faint radial image also demonstrates the power of IFU instruments for such work, especially of MUSE which combines a large redshift range with a large field of view and a high spatial resolution.

Regarding gas distribution in MACS\,J0417, we use \emph{Chandra} observations of our target to  identify an offset with respect to the light/dark-matter peak of $1.7\pm0.5$\arcsec. Following the method developed by Harvey et al.\ (\emph{in prep.}) we derive from this offset an upper limit to the self-interaction cross-section of dark matter of $\sigma_{\rm DM}/m < 3$\,cm$^2$.g$^{-1}$, in good agreement with previous measurements.
However, we note that the X-ray peak coincides with the optical emission-line peak of the BCG initially observed by \cite{green16}, and confirmed by our MUSE observations. Using this information, we estimate that the gas-dark matter-light offset can be explained by the on-going merger event in this actively evolving system \citep{ME12}: during the collision with the North-Western subcluster, the gaseous intra-cluster medium was held back while the effectively non-collisional dark matter and main stellar components continued unimpeded on their trajectory, and it is the local concentration of cold gas that led to direct star formation at a location offset from the light peak of the BCG.

We also compares our mass model with the one from \cite{mahler18b} and find excellent agreement in terms of rms (they obtain an rms of 0.37\arcsec) and total mass ($M({\rm R<200\,kpc})=(1.78\pm0.02)\times10^{14}\,\msun$). Since the two mass models were created independently, we investigate the differences between them in more detail by considering two different model outputs: (1) the two-dimensional density profile, and (2) the surface area in the source plane, $\sigma_{\mu}$, above a certain magnification $\mu$, which is a good estimator of the lensing power of a cluster.
The two density profiles show a slight discrepancy in the inner region, $R<10$\,kpc, which we attribute to the inclusion of the fourth image of system \#1 in our model. 
The lensing power of MACS\,J0417 predicted by both our \emph{Fiducial Model} and the model derived by \cite{mahler18b} is very similar; specifically, the two studies find $\sigma(\mu>3)=0.22$\,arcmin$^2$ and $\sigma(\mu>3)=0.23$\,arcmin$^2$, respectively. 
We note that these values are comparable to those previously measured by us for the HFF target MACS\,J0416.1-2403 \citep[$\sigma(\mu>3)=0.26$\,arcmin$^2$;][]{jauzac14}, underlining that MACS\,J0417 is indeed a powerful lens and well suited to probe the high-redshift universe \citep[see][for more details]{mahler18b}.

The scientific efficiency of MUSE observations of massive clusters is especially high, as they allow the spectroscopic identification and charaterization of both numerous cluster members and strongly lensed background galaxies, soon resulting in a sample of dozens of spectroscopically selected lensed galaxies for the community to follow up with new facilities such as ALMA, \emph{JWST} and SKA.

\section*{Acknowledgements}

The authors thank the anonymous referee for their useful comments.
MJ thanks D. Eckert for useful discussions.
We thank the RELICS team for releasing the fully reduced \emph{HST} imaging data, available to the community as hlsp through MAST. We thank Joshua Stephenson for his hard work creating the MUSE observation files.
This work was supported by the Science and Technology Facilities Council (grant numbers ST/L00075X/1, ST/P00541/1) and used the DiRAC Data Centric system at Durham University, operated by the Institute for Computational Cosmology on behalf of the STFC DiRAC HPC Facility (\url{www.dirac.ac.uk}). This equipment was funded by BIS National E-infrastructure capital grant ST/K00042X/1, STFC capital grant ST/H008519/1, and STFC DiRAC Operations grant ST/K003267/1 and Durham University. DiRAC is part of the National E-Infrastructure.
This project has received funding from the European Research Council (ERC) under the European Union's Horizon 2020 research and innovation programme (grant agreement No 757535).
This paper is based on observations made with the NASA/ESA \emph{Hubble Space Telescope}, obtained at the Space Telescope Science Institute, which is operated by the Association of Universities for Research in Astronomy, Inc., under NASA contract NAS 5-26555. These observations are associated with program GO-14096. Archival data are associated with program GO-12009.



\bibliographystyle{mnras}
\bibliography{reference} 



\appendix
\label{sec:appendix}

\begin{center}
\begin{figure*}
\includegraphics[width=0.245\textwidth,angle=0.0]{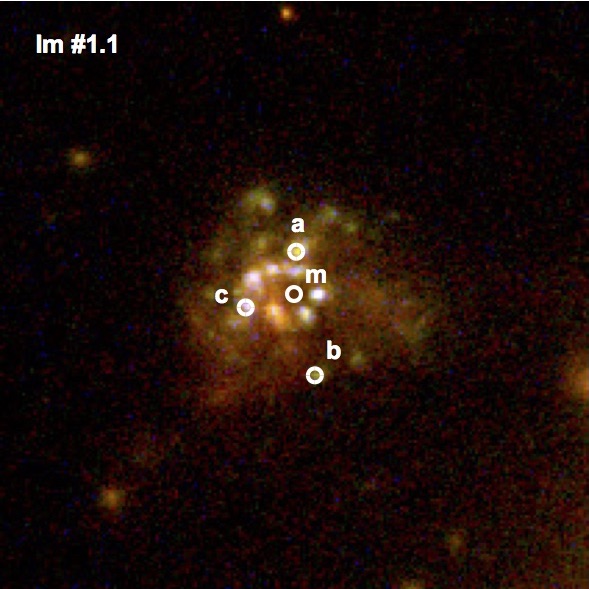}\
\includegraphics[width=0.245\textwidth,angle=0.0]{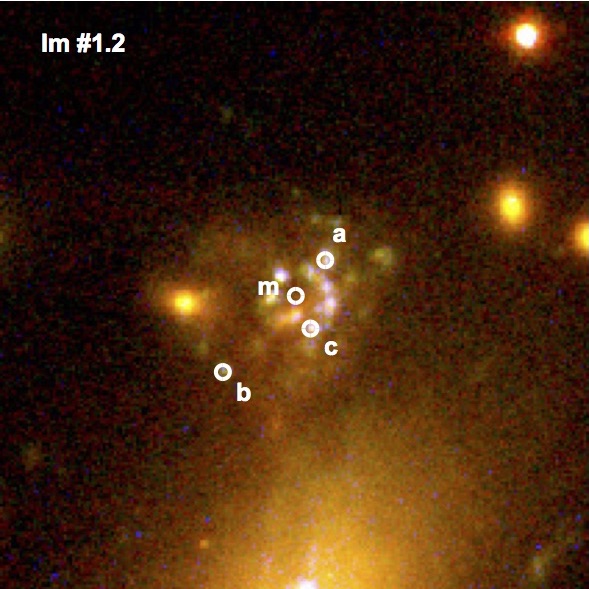}\
\includegraphics[width=0.245\textwidth,angle=0.0]{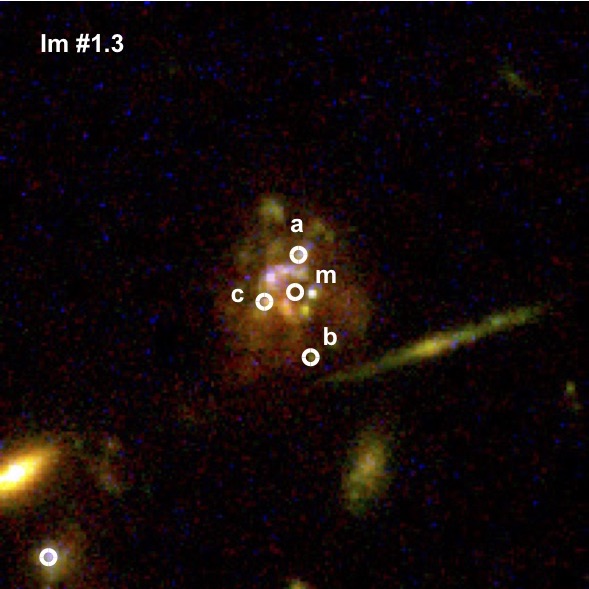}\
\includegraphics[width=0.245\textwidth,angle=0.0]{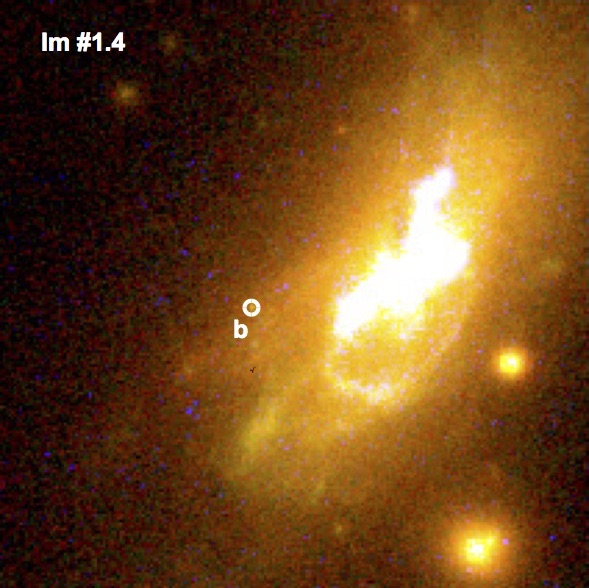}\
\\
\includegraphics[width=0.245\textwidth,angle=0.0]{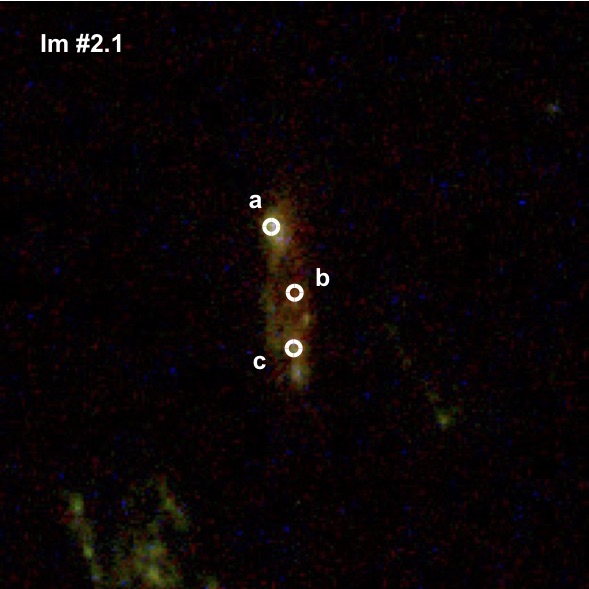}\ 
\includegraphics[width=0.245\textwidth,angle=0.0]{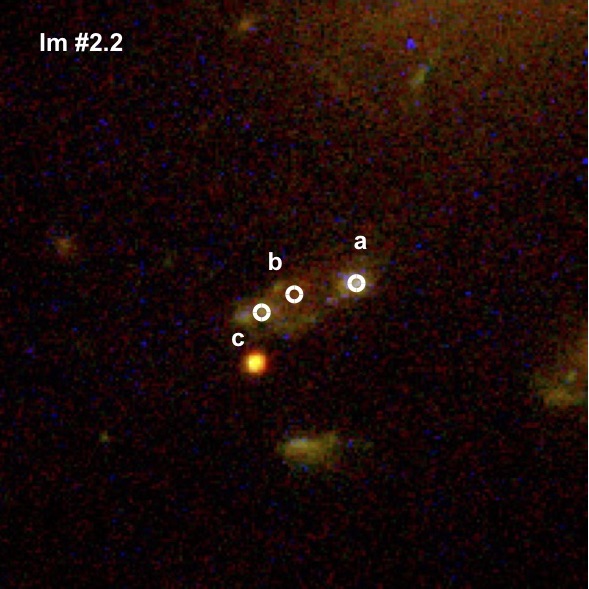}\ 
\includegraphics[width=0.245\textwidth,angle=0.0]{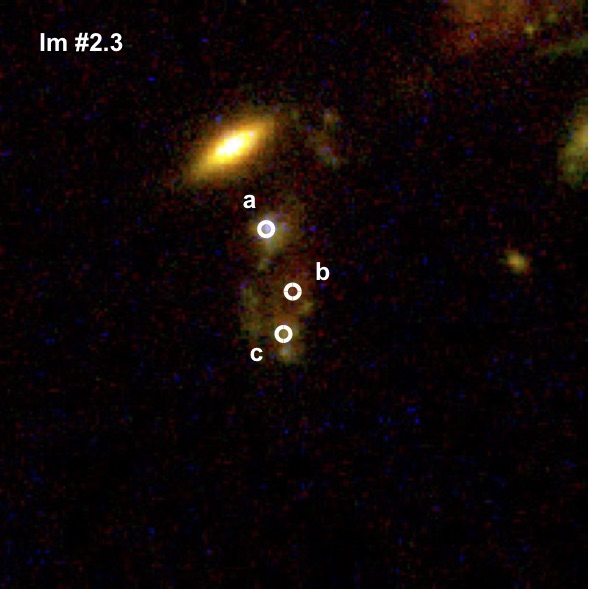}\
\includegraphics[width=0.245\textwidth,angle=0.0]{empty.pdf}\
\\
\caption{\emph{HST} colour composite stamps of the four multiple images of system \#1 (\emph{Top panel}), and the three multiple images of system \#2 (\emph{Bottom panel}). White circles highlight the position of the star formations knots used in the mass modeling, as listed in Table~\ref{tab:immul}. The size of the stamps is the same as in Fig.~\ref{fig:donut}.
}
\label{fig:sfknots}
\end{figure*}
\end{center}

\begin{center}
\begin{figure*}
\includegraphics[width=0.33\textwidth,angle=0.0]{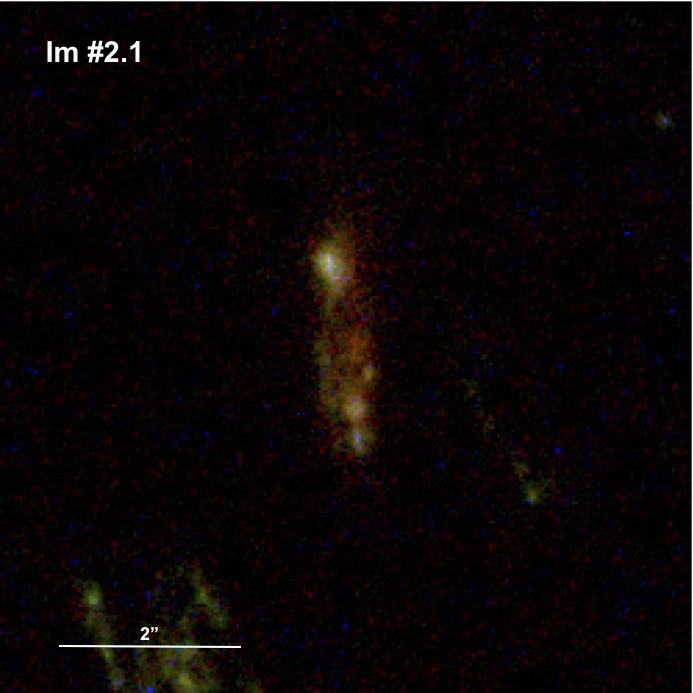}\
\includegraphics[width=0.33\textwidth,angle=0.0]{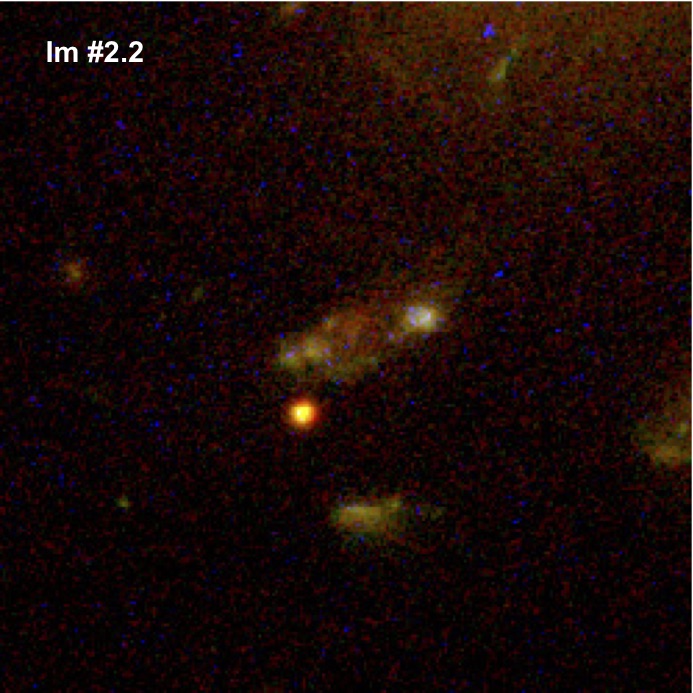}\
\includegraphics[width=0.33\textwidth,angle=0.0]{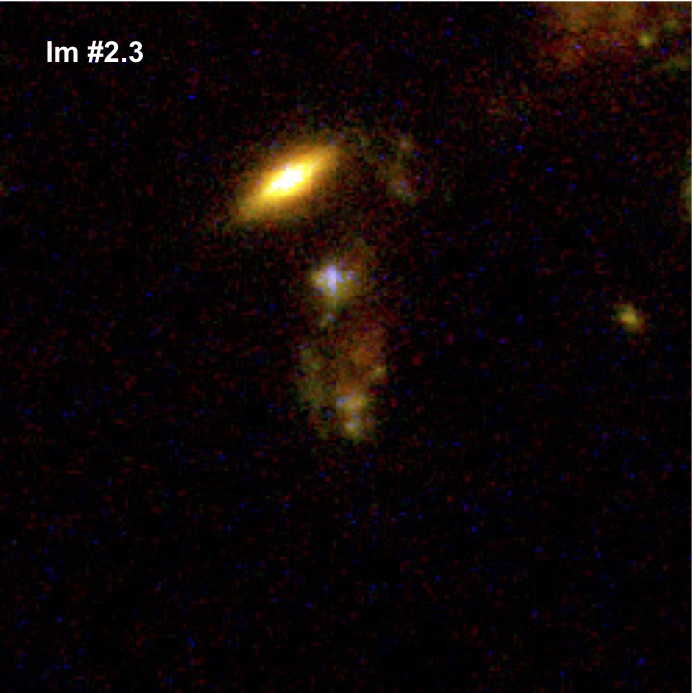}\
\\
\includegraphics[width=0.33\textwidth,angle=0.0]{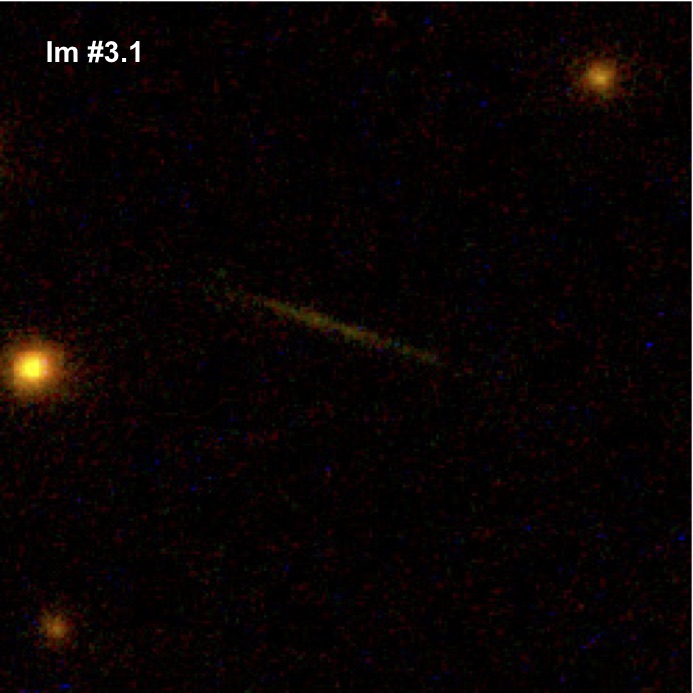}\ 
\includegraphics[width=0.33\textwidth,angle=0.0]{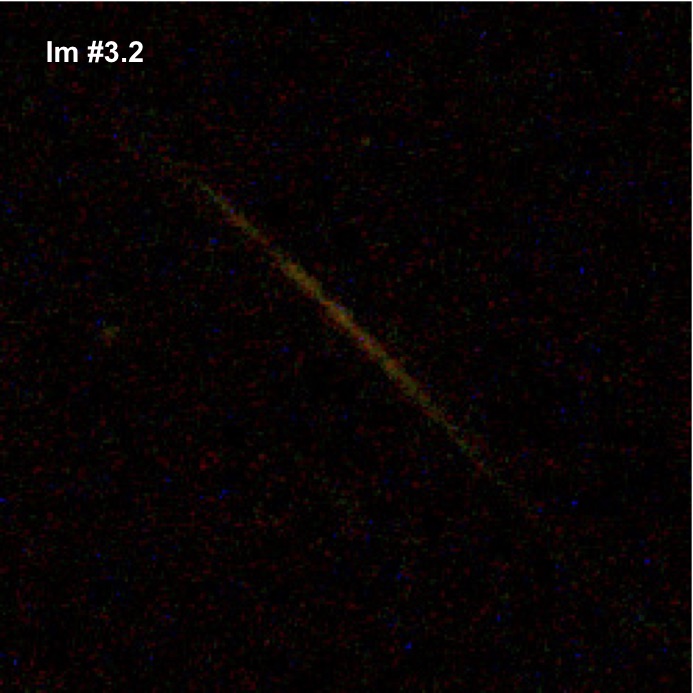}\ 
\includegraphics[width=0.33\textwidth,angle=0.0]{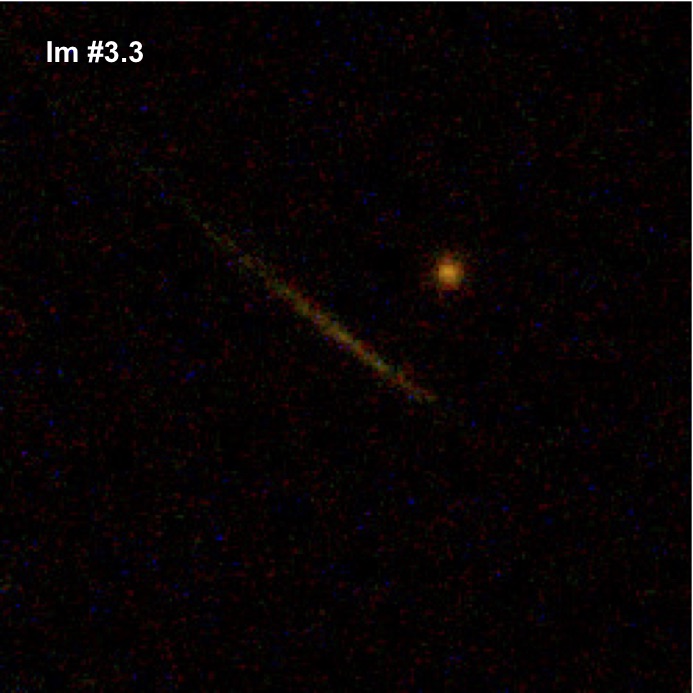}\ 
\\
\includegraphics[width=0.33\textwidth,angle=0.0]{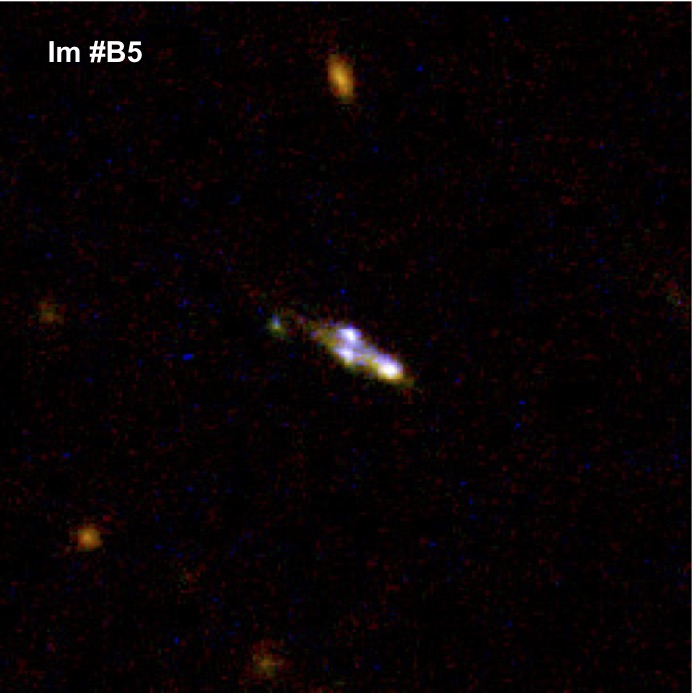}\ 
\includegraphics[width=0.33\textwidth,angle=0.0]{empty.pdf}\ 
\includegraphics[width=0.33\textwidth,angle=0.0]{empty.pdf}\ 
\\
\caption{Galaxy triplet at $z=1.046$. The three stamps of the first and second raws show the three lensed images of system \#2 and system \#3 respectively from the \emph{HST} colour composite image.
The third row shows the third galaxy at $=1.046$ identified in the MUSE data cube, which is strongly lensed but not multiply-imaged as it appears in the \emph{HST} colour composite image, the second stamp shows the source reconstruction.
Source \#2 and \#3 are separated by 130.4\,kpc, source \#2 and source \#B5 are separated by 154.1\,kpc, and source \#3 and source \#B5 are separated by 284.3\,kpc. Such separations imply this galaxy triplet is not interacting.
}
\label{fig:gal_trip}
\end{figure*}
\end{center}

\section{Multiple images of systems \#1 \& \#2}
\label{sec:sfknots}
Figure~\ref{fig:sfknots} shows the different components in each multiple images of system \#1 and system \#2 used as constraints in our mass model presented in Table~\ref{tab:immul}.

\section{A galaxy triplet at $z=1.046$}
\label{sec:galpair}
MUSE observations allowed us to measure the spectroscopic redshift of three multiply-imaged systems, two of which, system \#1 and system \#2, are at $z=1.046$. Adding to that, a third galaxy at $z=1.046$ is detected. This one is located outside the multiple image region, it is still relatively highly magnified, $\mu=6.2\pm2.0$, but not multiply imaged. 
As our mass model allows us to retrace back into the source plane, we can thus investigate this coincidence of three galaxies at the same redshift, and for example identify any possible interactions. 
%
Fig.~\ref{fig:gal_trip} shows the \emph{HST} stamps of the three images of systems \#2 and system \#3, as well as of galaxy \#B5. 

From the lens model we can measure the position of all three galaxies in the source plane at $z=1.046$, and thus measure the distance between them. We give those distances both in arcseconds and kpc in Table~\ref{tab:galtrip}. The multiply-imaged galaxies, galaxies \#2 and \#3 in Table~\ref{tab:galtrip}, are separated by 130.4\,kpc, while galaxy \#2 is located at 154.1\,kpc from galaxy \#B5. Such large distances imply those three galaxies are not interacting at the moment, however they could be part of a background group or cluster at $z=1.046$. Their large separation does not imply they did not interact in the past.

In Table~\ref{tab:galtrip} we also give the distances measured by Mahler et al. (2018) in kpc for comparison. Both our models predict distances that are in excellent agreement. This independent result reinforces the discussion in Sect.~\ref{sec:comp_muserel} which concludes both models are giving extremely similar results.

\begin{table}
\begin{center}
\caption{Distances between each galaxies part of the galaxy triplet detected in MUSE at $z=1.046$. At such redhsift, $1\arcsec=8.09$\,kpc.}
\begin{tabular}[h!]{cccc}
\hline
\hline
\noalign{\smallskip}
 & $D_{\rm \#2-\#3}$ & $D_{\rm \#2-\#B5}$ & $D_{\rm \#3-\#B5}$\\
\hline
\hline
\emph{Fiducial Model} & 16.1\arcsec & 19.1\arcsec & 35.1\arcsec \\
 & 130.4\,kpc & 154.1\,kpc & 284.3\,kpc \\
\hline
\emph{Mahler et al.\ 2018} & 131\,kpc & 157\,kpc & 288.6\,kpc \\
\hline
\hline
\end{tabular}
\label{tab:galtrip}
\end{center}
\end{table}

\section{A dense starburst at $z=0.56$}
\label{sec:greenpea}
One object detected in the MUSE field that stood out was object \#B1 at $z=0.5635$ in Table~\ref{tab:bkg_muse} that is dominated
by narrow [OIII] 4959\,\AA\ and 5007\,\AA\ with no detectable [OII] and H$\beta$ emission. The very
high equivalent width of this source of $>150$ implies that it is an extreme starburst 
in the ``Green Pea'' class \citep{cardamone09}. These galaxies are rare as only two [OIII]-only
emission galaxies were found in the first MUSE survey field in the HUDF out of 1338 sources 
\citep{inami17}. The HST imaging shows a 
compact but extended object with an AB magnitude of around 25.8~mag in the F814W filter
so considering the likely line contamination in the band the stellar mass of the galaxy
is likely to be below $10^{9}\,\msun$. There is no significant
X-ray emission from the source in the {\it Chandra} observation so the X-ray luminosity of
the galaxy is $<2\times$10$^{42}$~erg~s$^{-1}$. It is also undetected in  SPIRE
at 250\,$\mu$m from the {\it Herschel} Lensing Survey-snapshot project \citep{egami10}
giving an upper limit on star formation of $\sim10\,\msun$\,yr$^{-1}$ so consistent
with the observed specific star formation rates measured for these remarkable galaxies.

The implied amplification from our \emph{Fiducial Model} for this galaxy, $\mu=1.4\pm0.4$, is relatively low but the short lens to object distance is the dominant factor in this.

Determining the wider spectral properties of this intriguing source in the UV and NIR
would be straightforward given the nature of the line emission and our larger MUSE
survey and all other high Galactic latitude observations will reveal how unusual
this class of emission line galaxy is.


\bsp	
\label{lastpage}
\end{document}